\documentclass[a4paper,12pt]{article}
\usepackage[round]{natbib}
\usepackage{amssymb,lscape,psfrag}
\usepackage{epsfig,color}
\usepackage{array}
\usepackage[round]{natbib}
\usepackage{enumerate}
\usepackage{amsmath}
\usepackage{graphics}
\usepackage{graphicx}
\usepackage{epsfig}
\usepackage{verbatim}
\usepackage{cancel}
\usepackage{longtable}
\usepackage{multirow}
\usepackage{float}
\usepackage{ulem} 
\usepackage{epstopdf}

\begin{document}
\newcommand {\inispace}{\renewcommand{\baselinestretch}{1.}\normalsize}
\newcommand {\tabspace}{\renewcommand{\baselinestretch}{1.}\normalsize}
\newcommand{\thetab}{\boldsymbol\theta}
\newcommand{\phib}{\boldsymbol\phi}

\inispace

\title{On the use of marginal posteriors in marginal likelihood estimation via importance sampling}
\date{}
\author{
K. Perrakis\footnote{Department of Hygiene, Epidemiology and Medical Statistics, Medical School, University of Athens; kperrakis@med.uoa.gr},
I. Ntzoufras\footnote{Department of Statistics, Athens University of Economics and Business; ntzoufras@aueb.gr} 
and 
E.G. Tsionas\footnote{Department of Economics, Athens University of Economics and Business; tsionas@aueb.gr} 
}

\maketitle

\begin{abstract}
We investigate the efficiency of a marginal likelihood estimator where the product of the marginal posterior distributions is used as an importance sampling function.
The approach is generally applicable to multi-block parameter vector settings, does not require additional Markov Chain Monte Carlo (MCMC) sampling and is not dependent on the type of MCMC scheme used to sample from the posterior.
The proposed approach is applied to normal regression models, finite normal mixtures and longitudinal Poisson models, and leads to accurate marginal likelihood estimates. 

\vspace{0.2cm} \noindent \textit{Keywords:} Finite normal mixtures, importance sampling, marginal posterior, marginal likelihood estimation, random effect models, Rao-Blackwellization

\end{abstract}

\section{Introduction}

The problem of estimating the marginal likelihood has received considerable
attention during the last two decades. The topic is of importance
in Bayesian statistics as it is associated with the evaluation of
competing hypotheses or models via Bayes factors and posterior model
odds. Consider, briefly, two competing models $M_{1}$ and $M_{2}$
with corresponding prior probabilities $\pi(M_{1})$ and $\pi(M_{2})=1-\pi(M_{1})$.
After observing a data vector $\mathbf{y}$, the evidence in favour
of $M_{1}$ (or against $M_{2}$) is evaluated through the odds of
the posterior model probabilities $p(M_{1}|\mathbf{y})$ and $p(M_{2}|\mathbf{y})$,
that is, 
\[
\frac{p(M_{1}|\mathbf{y})}{p(M_{2}|\mathbf{y})}=\frac{m(\mathbf{y}|M_{1})}{m(\mathbf{y|}M_{2})}\times\frac{\pi(M_{1})}{\pi(M_{2})}.
\]
The quantity $B_{12}=m(\mathbf{y}|M_{1})/m(\mathbf{y}|M_{2})$ is
the ratio of the marginal likelihoods or prior predictive distributions
of $M_{1}$ and $M_{2}$ and is called the Bayes factor of $M_{1}$
versus $M_{2}$. The Bayes factor can also be interpreted as the ratio
of the posterior odds to the prior odds. When $M_{1}$ and $M_{2}$
are assumed to be equally probable a-priori, the Bayes factor is equal
to the posterior odds. 

The marginal likelihood of a given model $M_{k}$ associated with
a parameter vector $\boldsymbol{\theta}_{k}$ is essentially the normalizing
constant of the posterior $p(\boldsymbol{\theta}_{k}|\mathbf{y},M_{k})$,
obtained by integrating the likelihood function $l(\mathbf{y}|\boldsymbol{\theta}_{k},M_{k})$
with respect to the prior density $\pi(\boldsymbol{\theta}_{k}|M_{k})$,
i.e. 
\begin{equation}
m(\mathbf{y}|M_{k})=\int l(\mathbf{y}|\boldsymbol{\theta}_{k},M_{k})\pi(\boldsymbol{\theta}_{k}|M_{k})\mathrm{d}\boldsymbol{\theta}_{k}.
\end{equation}
The integration in (1) may be evaluated analytically for some elementary
cases. Most often, it is intractable thus giving rise to the marginal
likelihood estimation problem. Numerical integration methods can be
used as an approach to the problem, but such techniques are of limited
use when sample sizes are moderate to large or when vector $\boldsymbol{\theta}_{k}$
is of large dimensionality. 
In addition, the simplest Monte Carlo (MC) estimate, 
which is given by 
\begin{equation}
\widehat{m}(\mathbf{y}|M_{k})=N^{-1}\underset{n=1}{\overset{N}{\sum}}l(\mathbf{y}|\boldsymbol{\theta}_{k}^{(n)},M_{k}),
\end{equation} 
using draws $\{\boldsymbol{\theta}_{k}^{(n)}:n=1,2,...,N\}$ from the prior distribution  
is extremely unstable when the posterior is concentrated in relation to the prior. 
This scenario is frequently met in practice when  flat, low-information prior distributions are used to express prior ignorance. 
A detailed discussion regarding Bayes factors and marginal likelihood estimation is provided by Kass
and Raftery (1995).

It is worth noting that the problem of estimating (1) can be bypassed
by considering model indicators as unknown parameters. This option
has been investigated by several authors (e.g. Green, 1995, Carlin
and Chib, 1995, Dellaportas et al., 2002) who introduce MCMC algorithms
which sample simultaneously over parameter and model space and deliver
directly posterior model probabilities. 
However, implementation of these methods can get quite complex since they require enumeration
of all competing models and specification of tuning constants or ``pseudopriors''
(depending upon approach) in order to ensure successful mixing in model space. 
Moreover, since these methods focus on the estimation of posterior model probabilities, 
accurate estimation of the marginal likelihoods and/or Bayes factors will not be feasible in the 
cases where a dominating model exists in the set of models under consideration; 
tuning, following the lines of Ntzoufras et al. (2005),  might be possible but is typically inefficient and time consuming. 

In contrast, ``direct'' methods provide marginal likelihood estimates
by utilizing the posterior samples of separate models. These methods
are usually simpler to implement and are preferable in practice when
the number of models under consideration is not large, namely when
it is practically feasible to obtain a posterior sample for each
of the competing models. Work along these lines includes the Laplace-Metropolis
method (Lewis and Raftery, 1997), the harmonic-mean and the prior/posterior
mixture importance sampling estimators (Newton and Raftery, 1994), bridge-sampling methods (Meng
and Wong, 1994),
candidate's estimators for Gibbs sampling (Chib, 1995) and 
Metropolis-Hastings sampling (Chib and Jeliazkov, 2001), 
annealed importance sampling (Neal, 2001), 
importance-weighted marginal density estimators (Chen, 2005) and 
nested sampling approaches (Skilling, 2006). 
More recently, Raftery et al. (2007) presented a stabilized version of the harmonic-mean estimator, 
while Friel and Pettitt (2008) and Weinberger (2012) proposed new approaches based on power posteriors and Lebesgue
integration theory, respectively. 
It is worth mentioning that Bayesian evidence evaluation is also of particular interest in the astronomy literature where nested sampling is commonly used for marginal likelihood estimation (e.g. Feroz et al., 2009; Feroz et al., 2011). Recent reviews comparing popular methods based on MCMC sampling can be found in Friel and Wyse (2012) as well as in Ardia et al. (2012). Alternative approaches for marginal likelihood estimation include sequential Monte Carlo (Del Moral et al., 2006) and variational Bayes (Parise and Welling, 2007) methods.

In this paper we propose using the marginal posterior distributions on importance sampling estimators of the marginal likelihood. 
The proposed approach is particularly suited for the Gibbs sampler,
but it is also feasible to use for other types of MCMC algorithms. 
The estimator can be implemented in a straightforward manner and 
it can be extended to multi-block parameter settings without requiring additional MCMC sampling apart
from the one used to obtain the posterior sample. 

The remainder of the paper is organized as follows. The proposed estimator
and its variants are discussed in Section 2. In Section 3 the method
is applied to normal regression models, to finite normal mixtures
and also to hierarchical longitudinal Poisson models. Concluding remarks
are provided in Section 4.

\section{The proposed estimator}

In the following we first introduce the proposed estimator in a two block setting. 
The more general multi-block case is considered next, explaining why the estimator will be useful in such cases.  
We further present details concerning the implementation of the proposed approach when the model formulation 
includes latent variables or nuisance parameters that are not of prime interest for model inference. 
The section continues with a description of the different estimation approaches of the posterior marginal distributions 
used as importance functions. 
We conclude with a note on a convenient implementation of the estimator for models where the posterior distribution becomes invariant under competing diffuse priors and brief remarks about the 
calculation of numerical standard errors.
In the remaining of the paper, the dependence to the model indicator $M_k$ (introduced in the previous section) is eliminated 
for notational simplicity.

\subsection{Introducing the estimator in a two-block setting}

Let us consider initially the 2-block setting
where $l(\mathbf{y}|\boldsymbol{\theta},\boldsymbol{\phi})$
is the likelihood of the data conditional on parameter vectors $\boldsymbol{\theta}=(\theta_{1},\theta_{2},...,\theta_{p})^{T} $
and $\boldsymbol{\phi}=(\phi_{1},\phi_{2},...,\phi_{q})^{T}$, which
can be either independent, i.e. $\pi(\boldsymbol{\theta},\boldsymbol{\phi})=\pi(\boldsymbol{\theta})\pi(\boldsymbol{\phi})$,
or dependent, e.g. $\pi(\boldsymbol{\theta},\boldsymbol{\phi})=\pi(\boldsymbol{\theta}|\boldsymbol{\phi})\pi(\boldsymbol{\phi})$,
a-priori. In general, one can improve the estimator in (2) by introducing
a proper importance sampling density $g$ and then calculate the marginal
likelihood as an expectation with respect to $g$ instead of the prior,
i.e. 
\begin{equation*}
m(\mathbf{y})=\int\frac{l(\mathbf{y}|\boldsymbol{\theta},\boldsymbol{\phi})\pi(\boldsymbol{\theta},\boldsymbol{\phi})}{g(\boldsymbol{\theta},\boldsymbol{\phi})}g(\boldsymbol{\theta},\boldsymbol{\phi})\mathrm{d}(\boldsymbol{\theta},\boldsymbol{\phi})=\mathrm{E}_{g}\left[\frac{l(\mathbf{y}|\boldsymbol{\theta},\boldsymbol{\phi})\pi(\boldsymbol{\theta},\boldsymbol{\phi})}{g(\boldsymbol{\theta},\boldsymbol{\phi})}\right].
\end{equation*}
This quantity can be easily estimated as
\begin{equation*}
\widehat{m}(\mathbf{y})=N^{-1}\underset{n=1}{\overset{N}{\sum}}\frac{l(\mathbf{y}|\boldsymbol{\theta}^{(n)},\boldsymbol{\phi}^{(n)})\pi(\boldsymbol{\theta}^{(n)},\boldsymbol{\phi}^{(n)})}{g(\boldsymbol{\theta}^{(n)},\boldsymbol{\phi}^{(n)})},
\end{equation*}
where $\boldsymbol{\theta}^{(n)}$ and $\boldsymbol{\phi}^{(n)}$,
for $n=1,2,...,N$, are draws from $g$. Theoretically, an ideal importance
sampling density is proportional to the posterior. In practice, we
seek densities which are similar to the posterior and easy to sample
from. 

Given this consideration, we propose to use the product of the marginal
posterior distributions as importance sampling density, i.e. 
$g(\boldsymbol{\theta},\boldsymbol{\phi}) \equiv p(\boldsymbol{\theta}|\mathbf{y})p(\boldsymbol{\phi}|\mathbf{y})$.
Under
this approach
\begin{equation*}
m(\mathbf{y})= \int \int\frac{l(\mathbf{y}|\boldsymbol{\theta},\boldsymbol{\phi})\pi(\boldsymbol{\theta},\boldsymbol{\phi})}{p(\boldsymbol{\theta}|\mathbf{y})p(\boldsymbol{\phi}|\mathbf{y})}p(\boldsymbol{\theta}|\mathbf{y})p(\boldsymbol{\phi}|\mathbf{y})\mathrm{d}\boldsymbol{\theta}\mathrm{d}\boldsymbol{\phi},
\end{equation*}
which yields the estimator
\begin{equation}
\widehat{m}(\mathbf{y})=N^{-1}\underset{n=1}{\overset{N}{\sum}}\frac{l(\mathbf{y}|\boldsymbol{\theta}^{(n)},\boldsymbol{\phi}^{(n)})\pi(\boldsymbol{\theta}^{(n)},\boldsymbol{\phi}^{(n)})}{p(\boldsymbol{\theta}^{(n)}|\mathbf{y})p(\boldsymbol{\phi}^{(n)}|\mathbf{y})}.
\end{equation}
Note that the only twist in (3) is that the draws $\boldsymbol{\theta}^{(n)}$
and $\boldsymbol{\phi}^{(n)}$, for $n=1,2,...,N$, are draws
from the marginal posteriors $p(\boldsymbol{\theta}|\mathbf{y})$
and $p(\boldsymbol{\phi}|\mathbf{y})$ and not from the joint posterior
$p(\boldsymbol{\theta},\boldsymbol{\phi}|\mathbf{y})$. In
most cases the marginal posterior distributions will not be known,
nevertheless, this does not constitute a major obstacle neither for
sampling from the marginal posteriors nor for calculating the marginal
probabilities which appear in the denominator of (3); the former issue is discussed here, the latter
is handled in Section 2.4. 

It is straightforward to see that the product marginal posterior is the optimal importance sampling density when $\boldsymbol\theta$ and $\boldsymbol\phi$ are independent a-posteriori, since in this case $p(\boldsymbol\theta|y)p(\boldsymbol\phi|y)=p(\boldsymbol{\theta},\boldsymbol{\phi}|y)$ leading to the zero-variance estimator. Although posterior independence is not frequently met in practice, the product marginal posterior can serve as a good approximation to the joint posterior even if $\boldsymbol{\theta}$ and $\boldsymbol{\phi}$ are not completely independent a-posteriori. 
First, it has exactly the same support as the joint posterior. Second, the blocking of the parameters can be such that the parameter blocks are close to orthogonal regardless whether the elements within $\thetab$ and $\phib$ are strongly correlated.
Furthermore, appropriate reparameterizations can be used in order to form parameter blocks which are orthogonal or close to orthogonal (see e.g. Gilks and Roberts, 1996). 
Moreover, in generalized linear models the augmentation scheme of Ghosh and Clyde (2011)  can be used to obtain orthogonal parameters. 

It is worth noting that the estimator in (3) is similar to the Markov chain importance sampling approach described in Botev et al. (2012) and the marginal likelihood estimator proposed in Chan and Eisenstat (2013) based on the cross-entropy method. Botev et al. (2012) show that the product of the marginals is the best importance sampling density -- in the sense of minimizing the Kullback-Leibler divergence with respect to the zero-variance importance sampling density -- among all product form importance sampling densities, given that the zero-variance density is also decomposable in product form. Similarly, Chan and Eisenstat (2013) locate the importance sampling density minimizing Kullback-Leibler divergence with respect to block-independent factorizations of the joint posterior from distributions belonging to the same parametric families as the priors. 
The approach presented here differentiates from the above estimators 
since we consider directly the marginal posteriors and ``manipulate'' the joint MCMC sample in order to construct marginal samples, thus avoiding further importance sampling and leaving estimation of marginal densities as the main issue to deal with.

In general, marginal posterior samples can be obtained from any MCMC algorithm. 
The only problem is that a single MCMC chain corresponds to 
a sample from the joint distribution, with non-zero covariance between parameter blocks. 
One option is to use a different MCMC run for each block of parameters. In this case,
one can calculate the estimator in (3) by using draws  $\boldsymbol{\theta}^{(n)}$, $\boldsymbol{\phi}^{(n)}$
coming from two independent MCMC samples of equal size $N$.  
Nevertheless, this approach can considerably increase the number of MCMC iterations, 
especially for a large number of parameter blocks. In addition, the approach is not economical
in the sense that only $N$ posterior draws are used from a total sample of $2N$ draws.

A simple and more efficient solution is to re-order a single MCMC chain in such a way that
it does not correspond to a sample from the joint posterior distribution. 
This can be easily implemented by systematically permuting either the sampled values of $\boldsymbol{\theta}$
or those of $\boldsymbol{\phi}$. For instance, consider one MCMC chain  
where
the initial MCMC draws are indexed as $\{\boldsymbol{\theta}^{(n)},\boldsymbol{\phi}^{(n)}:n=1,2,...,N_{1},N_{1}+1,N_{1}+2,...,N\}$
where $N_{1}=N/2$. Then, one can simply re-order the sample of $\boldsymbol{\phi}$
as $\{\boldsymbol{\phi}^{(n_{1})}:n_{1}=N_{1}+1,N_{1}+2,...,N,1,2,,...,N_{1}\}$ and join the set of draws $\{\boldsymbol{\mathbf{\mathbf{\theta}}}^{(n)},\mathbf{\boldsymbol{\phi}}^{(n_{1})}\}$,
thus forming a sample of paired realizations from the 
distribution $g(\boldsymbol{\theta}, \boldsymbol{\phi})=p(\boldsymbol{\theta}|\mathbf{y}) p(\boldsymbol{\phi}|\mathbf{y})$.
Obviously, when the paired
sample has been formed, the distinction between the two sets of indices
becomes irrelevant and a common index may be adopted as presented
in equation (3). 
Reordering is trivial in implementation regardless
of the number of parameter blocks; the only initial requirement is
that the size of the final MCMC sample $N$ must be dividable with
the number of blocks, say $B$, so that $B$ independent reorderings of the
MCMC chain can be formed. This line of reasoning also holds for cases of multiple-chain MCMC sampling,   
which is a frequent MCMC strategy favoured mainly on the basis of MCMC convergence checks (e.g. Gelman and Rubin, 1992). 
For such cases, one can re-order 
the within-chain posterior samples in a similar manner to that previously described, and then form a joined sample from the multiple re-ordered chains.

Another, even simpler, alternative for removing correlations between marginal samples is to randomly permute each sample.
Results using this approach will be similar to the above systematic re-ordering, except in extreme, unlikely cases 
where randomly permuted samples with non-zero sample correlation are generated by chance. Irrespective of the reorderding scheme used, for the remainder of this paper we use a
common index $n$ across parameter blocks, referring to joined independent
block samples.

\subsection{Extension to multi-block settings}

Generalization to multi-block hierarchical settings is straightforward.
Consider $B$ blocks $\boldsymbol{\theta}_{1},\boldsymbol{\theta}_{2},...,\boldsymbol{\theta}_{B}$;
in this case the product of the $B$ marginal posteriors is used as
importance sampling density and 
\[
m(\mathbf{y})= \int ...  \int\frac{l(\mathbf{y}|\boldsymbol{\theta}_{1},...,\boldsymbol{\theta}_{B})\pi(\boldsymbol{\theta}_{1},...,\boldsymbol{\theta}_{B})}{\underset{i=1}{\overset{B}{\prod}}p(\boldsymbol{\theta}_{i}|\mathbf{y})}\prod_{i=1}^{B} p(\boldsymbol{\theta}_{i}|\mathbf{y})\mathrm{d}\boldsymbol{\theta}_{i} ,
\]
which yields the following estimator
\begin{equation}
\widehat{m}(\mathbf{y})=N^{-1}\sum_{n=1}^{N}\frac{l(\mathbf{y}|\boldsymbol{\theta}_{1}^{(n)},...,\boldsymbol{\theta}_{B}^{(n)})\pi(\boldsymbol{\theta}_{1}^{(n)},...,\boldsymbol{\theta}_{B}^{(n)})}{\underset{i=1}{\overset{B}{\prod}}p(\boldsymbol{\theta}_{i}^{(n)}|\mathbf{y})}.
\end{equation}

Note that the estimator in (4) may also refer to multiple unidimensional blocks where 
each parameter forms one block. 
The advantage of such an estimator raises from the fact that it is easy to construct good approximations 
of univariate marginal posterior distributions. 
On the other hand, any possible gain in the efficiency earned from the construction of good approximating densities for the marginal posteriors 
might be moderated by the use of an importance function which assumes overall independency. 
Therefore, the most efficient strategy is to choose blocks of minimal size constituted only by highly correlated parameters, which have at the same time 
weak between-block corrrelations.
Of course, quite often the model design is such that this condition is already met, e.g. for reasons of efficient MCMC mixing.
In addition, for cases of Gibbs sampling the natural blocking is the most convenient to use,
since in such cases marginal posterior densities can be estimated accurately even for high-dimensional parameter blocks 
(see Section 2.4 for more details).

\subsection{Handling latent variables and nuisance parameters} 

Many hierarchical models include a block component $\mathbf{u}$,
usually not of main inferential interest, which is associated with a hyperparameter
vector $\boldsymbol{\omega}$ through the relationship $\pi(\mathbf{u},\boldsymbol{\omega})=\pi(\mathbf{u}|\boldsymbol{\omega})\pi(\boldsymbol{\omega})$.
For instance, $\mathbf{u}$ may be a random effect vector or a latent
vector used to facilitate posterior simulation through Gibbs sampling
as in the data augmentation setting introduced in Tanner and Wong (1987). 
In such cases inference usually focuses on the marginal sampling likelihood by integrating out $\mathbf{u}$. 
For example, when there
is only one parameter block $\boldsymbol{\theta}$, the marginal sampling
likelihood is $l(\mathbf{y}|\boldsymbol{\theta},\boldsymbol{\omega})=\int l(\mathbf{y}|\boldsymbol{\theta},\mathbf{u})\pi(\mathbf{u}|\boldsymbol{\omega})d\mathbf{u}$.
In this case, a marginal likelihood estimate is obtained through equation
(3), where $\boldsymbol{\phi}$ is simply replaced by $\boldsymbol{\omega}$.
The extension to the multi-block setting is essentially the same as
in (4), only with the addition of $\boldsymbol{\omega}$, specifically
\begin{equation}
\widehat{m}(\mathbf{y})=N^{-1}\sum_{n=1}^{N}\frac{l(\mathbf{y}|\boldsymbol{\theta}_{1}^{(n)},...,\boldsymbol{\theta}_{B}^{(n)},\boldsymbol{\omega}^{(n)})\pi(\boldsymbol{\theta}_{1}^{(n)},...,\boldsymbol{\theta}_{B}^{(n)})\pi(\boldsymbol{\omega}^{(n)})}{\underset{i=1}{\overset{B}{\prod}}p(\boldsymbol{\mathbf{\theta}}_{i}^{(n)}|\mathbf{y})p(\boldsymbol{\omega}^{(n)}|\mathbf{y})}.
\end{equation}
Alternatively, there is also the option of working with the hierarchical
likelihood and including $\mathbf{u}$ in the estimation process,
i.e.
\begin{equation}
\widehat{m}(\mathbf{y})=N^{-1}\sum_{n=1}^{N}\frac{l(\mathbf{y}|\boldsymbol{\theta}_{1}^{(n)},...,\boldsymbol{\theta}_{B}^{(n)},\mathbf{u}^{(n)})\pi(\boldsymbol{\theta}_{1}^{(n)},...,\boldsymbol{\theta}_{B}^{(n)})\pi(\mathbf{u}^{(n)}|\boldsymbol{\omega}^{(n)})\pi(\boldsymbol{\omega}^{(n)})}{\underset{i=1}{\overset{B}{\prod}}p(\boldsymbol{\theta}_{i}^{(n)}|\mathbf{y})p(\mathbf{u}^{(n)}|\mathbf{y})p(\boldsymbol{\omega}^{(n)}|\mathbf{y})}.
\end{equation}
The latter approach is less practical to implement as it requires evaluation of $p(\mathbf{u}|\mathbf{y})$. 
In addition, marginalization over $\mathbf u$ will in general lead to more precise marginal likelihood estimates due to scaling down the parameter space; see Vitoratou {\it et al.} (2013) for further details. 
Therefore, estimator (5) is overall preferable to estimator (6), except perhaps in cases where 
the likelihood in (5) is not available analytically and also
estimation of $p(\mathbf{u}|\mathbf{y})$ is easy to handle
based on the methods discussed next.

\subsection{Estimating marginal posterior densities}

As seen so far, the proposed approach is fairly simple to implement. The only remaining issue is
the evaluation of the marginal posterior probabilities appearing in the denominators of estimators (3)--(6).  
Here we discuss some different approaches that can be adopted. 

A first simple approach is to assume normality either directly or indirectly. Let us consider, for instance, the 2-block
setting of Section 2.1 with parameter blocks $\boldsymbol{\theta}$ and $\boldsymbol{\phi}$.
Suppose, for instance, that $\boldsymbol{\theta}$ relates to a vector of means or a vector of regression
parameters. Then, for moderate to large sample sizes, 
a reasonable option is to assume that
$\boldsymbol{\theta}|\mathbf{y}\sim\mathbf{N}(\overline{\boldsymbol{\theta}},\mathbf{\boldsymbol{\Sigma_{\theta}}})$,
where $\overline{\boldsymbol{\theta}}$ and $\boldsymbol{\Sigma_{\theta}}$
are the estimated posterior mean vector and variance-covariance matrix
from the MCMC output, respectively.
Vector $\boldsymbol{\phi}$, on the hand, may refer to a vector of disperion parameters,
where the assumption of normality may not be suitable. 
One stategy, often sufficient in many cases, is to 
assume that a transformation $\mathbf{x}=t(\boldsymbol{\phi})$
is approximately
normal, i.e. $\mathbf{x}|\mathbf{y}\sim\mathbf{N}(\overline{\mathbf{x}},\boldsymbol{\Sigma}_{\mathbf{x}})$,
and consequently $p(\boldsymbol{\phi}|\mathbf{y})=p(\mathbf{x}|\mathbf{y})|\frac{\mathrm{d}t^{-1}(\mathbf{x})}{\mathrm{d}\mathbf{x}}|^{-1}$
for an appropriate invertible function $t(\cdot)$.

An alternative is to mimic the marginal posteriors by adopting appropriate
distributional assumptions and matching parameters to posterior moments.
This option is more suitable when the assumption of normality is not 
particularly supported and appropriate transformation functions are hard to find. 
In such cases, one can also consider a wide range of options
based on multivariate kernel methods (e.g. Scott, 1992) as an efficient
alternative.

Moreover, when implementing Gibbs sampling 
where the normalizing constants of the full conditional distributions are known, 
marginal posterior densities can be estimated through 
an efficient, simulation-consistent technique referred as
Rao--Blackwellization by Gelfand and Smith (1990). Consider, for instance, the $B$ parameter block setting of Section 2.2; 
the Rao-Blackwell estimates in this case are 
\begin{eqnarray*} 
\widehat{p}(\boldsymbol{\theta}_1|\mathbf{y})
&=&L^{-1}\sum_{l=1}^{L}p(\boldsymbol{\theta}_1| 
                   \boldsymbol{\theta}^{(l)}_2, \dots, \boldsymbol{\theta}^{(l)}_B, \mathbf{y}), \\  
\widehat{p}(\boldsymbol{\theta}_b|\mathbf{y})
&=&L^{-1}\sum_{l=1}^{L}p(\boldsymbol{\theta}_b|
                   \boldsymbol{\theta}^{(l)}_1, \dots, \boldsymbol{\theta}^{(l)}_{b-1},
                   \boldsymbol{\theta}^{(l)}_{b+1}, \dots, \boldsymbol{\theta}^{(l)}_B , \mathbf{y}) 
                   \mbox{~for~} b=2,\dots,B-1 \\  
\widehat{p}(\boldsymbol{\theta}_B|\mathbf{y})
&=&L^{-1}\sum_{l=1}^{L}p(\boldsymbol{\theta}_B|
                   \boldsymbol{\theta}^{(l)}_1, \dots, \boldsymbol{\theta}^{(l)}_{B-1}, \mathbf{y}).
\end{eqnarray*}                      
Note that not all $N$ posterior draws need to be used; usually a sufficiently large
subsample of $L$ posterior draws is adequate. For instance, in the
examples presented next we find that samples between 200 to 500 draws
are sufficient, which significantly reduces computational expense. 
It should also be noted that Rao-Blackwell estimates must be based on draws from the joint posterior distribution,
that is draws from the initial non-permuted MCMC sample.

Finally, for cases of hybrid Gibbs sampling
where only some full conditionals are known,
one can use a combination of the methods discussed here.
Rao-Blackwellization may be used for parameter blocks with known full conditional distributions,
whereas for the remaining blocks one can choose among 
distributional approximations based on moment-fitting and kernel methods. 

Estimators (3)--(6) will not be unbiased when approximating the marginal posterior densities using moment-matching strategies.  
Nevertheless, in practice such ``proxies'' can be very accurate for univariate as well as multivariate distributions. 
For instance, as illustrated in Section 3.3, high-dimensional marginal posteriors are approximated efficiently through multivariate normal distributions. 
In addition, the degree of bias can be empirically checked by comparing such estimates to the corresponding ones 
using importance samples from the moment-matched approximating distributions. 
The latter procedure yields an unbiased estimator of the marginal likelihood and, therefore, 
small observed differences will imply that the bias introduced is negligible.

\subsection{Marginal likelihood estimation for diffuse priors}

As known, the marginal likelihood is very sensitive to changes in the prior distribution, 
whereas the posterior distribution (after a point) 
is insensitive to the prior as the latter becomes more and more diffuse. 
Therefore, a usual drawback of marginal likelihood estimators that are based solely on draws from the posterior distribution is that they are typically not reliable for evaluating the marginal likelihoods of different models when considering diffuse priors (see e.g. Friel and Wyse, 2012).

Nevertheless, this is not the case for the proposed estimator as it incorporates the prior in the estimation of the marginal likelihood. 
In fact, we can easily adopt estimator (3) in order to estimate the marginal likelihood under 
different diffuse priors (that have no essential effect on the posterior distribution) 
using a sample from a single MCMC run. 
To illustrate this, consider the 2-block setting of Section 2.1 and two diffuse priors $\pi_0, \pi_1$ under which the posterior distribution remains unchanged, i.e. $p_0 \equiv p_1$. 
Let us assume that draws  $\{\boldsymbol{\theta}^{(n)},\boldsymbol{\phi}^{(n)}:n=1,2,...,N\}$ are available from an initial MCMC run and that the marginal likelihood $m_0$ under $\pi_0$ has already been estimated through (3). Then, the marginal likelihood under $\pi_1$ can be accurately estimated by 
\begin{equation}
\widehat{m}_{1}(\mathbf{y})=N^{-1}\underset{n=1}{\overset{N}{\sum}}\frac{l(\mathbf{y}|\boldsymbol{\theta}^{(n)},\boldsymbol{\phi}^{(n)})\pi_{1}(\boldsymbol{\theta}^{(n)},\boldsymbol{\phi}^{(n)})}{p_{0}(\boldsymbol{\theta}^{(n)}|\mathbf{y})p_{0}(\boldsymbol{\phi}^{(n)}|\mathbf{y})}.
\end{equation}          
The estimator in (7) does not require additional MCMC sampling, likelihood evaluations or 
evaluations of the marginal posterior densities since the posterior distributions $p_1$ and $p_0$ 
are the same under $\pi_1$ and $\pi_0$, respectively; 
the only extra effort involved is calculation of the prior probabilities $\pi_{1}(\thetab^{(n)},\phib^{(n)})$, for $n=1,2,...,N$.

\subsection{Calculating the numerical standard error}
  
The method of batch means provides a straightforward way for calculating the numerical or MC error of the estimator. Consider for instance the 2-block setting of Section 2.1; in this case the block-independent posterior sample $\boldsymbol{\zeta}^{(n)}\equiv (\boldsymbol{\theta}^{(n)},\boldsymbol{\phi}^{(n)})$ is divided into $K$ batches $\boldsymbol{\zeta}_{1}^{(n_k)},\boldsymbol{\zeta}_{2}^{(n_k)},...,\boldsymbol{\zeta}_{K}^{(n_k)}$ of size $N_K$, i.e. $n_{k}=1,2,...,N_K$ and $N=KN_K$, and one calculates 
\begin{equation}
\widehat{m}_{k}(\mathbf{y})=N_{K}^{-1}\underset{n_{k}=1}{\overset{N_K}{\sum}}\frac{l(\mathbf{y}|\boldsymbol{\theta}_{k}^{(n_{k})},\boldsymbol{\phi}_{k}^{(n_{k})})\pi(\boldsymbol{\theta}_{k}^{(n_{k})},\boldsymbol{\phi}_{k}^{(n_{k})})}{p(\boldsymbol{\theta}_{k}^{(n_{k})}|\mathbf{y})p(\boldsymbol{\phi}_{k}^{(n_{k})}|\mathbf{y})},
\end{equation}
for $k=1,2,...,K$. Then, an estimate of the standard error is given by
\begin{equation}
\widehat{\mathrm{s.e.}}\left(\widehat{m}(\mathbf y)\right)=\sqrt{\frac{1}{K(K-1)}\underset{k=1}{\overset{K}{\sum}}\left[ \widehat{m}_{k}(\mathbf{y}) - \overline{m}(\mathbf{y})\right]^2 },
\end{equation} 
where $\overline{m}(\mathbf{y})=K^{-1}\sum_{k=1}^{K}\widehat{m}_{k}(\mathbf{y})$ is the average batch mean estimate. Note that $K$ must be large enough to ensure proper estimation of the variance (the usual choice is $30 \leq K \leq 50$) and $N_K$ must also be sufficiently large so that the $\widehat{m}_{k}$'s are roughly independent (see e.g. Carlin and Louis, 1996).

Alternatively, we can consider the variance estimators of Newey and West (1987) and Geyer (1992) for dependent MCMC draws. Such estimators are suited when systematic re-ordering is used to form the block-independent posterior sample, since, in this case, the posterior dependency patterns will be the same as those of the initial MCMC sample. This is due to the fact that the number of parameter blocks $B$ will be usually much smaller 
than the size of the posterior sample ($B\ll N$). Therefore, serial auto-correlations for lags greater than $N/B$ are expected to be negligible (for converged MCMC runs), while auto-correlations of lower order are not affected by the re-ordering.

Finally, checking whether the variance is finite or not can be investigated empirically; if the variance is finite then one should expect that increasing the MCMC sample by a factor of $d$ should lead to a decrease of the standard error estimate by a factor approximately equal to $\sqrt d$. 
As illustrated in Section 3.4, the variance of the proposed estimator is finite for the examples presented in Section 3.

\section{Examples}

In this section we apply our method to three common classes of models.
First, we consider normal linear regression where the true marginal
likelihood can be calculated analytically, and compare the proposed
estimator to other estimators commonly used in practice. The second
example concerns finite normal mixture models where marginal likelihood
estimation has proven particularly problematic due to non-identifiability. 
In the third example, we apply the proposed methods to an hierarchical longitudinal Poisson model 
where the integrated sampling likelihood is analytically unavailable and, furthermore, 
standard Gibbs sampling cannot be implemented. 
The section closes with an empirical diagnostic for checking the assumption of finite variance by comparing the corresponding errors from samples of size $N$ and $2N$.

In all illustrations, we denote the likelihood 
functions with $l(\cdot)$, prior densities with $\pi(\cdot)$ and
posterior or full conditional distributions with $p(\cdot)$. Concerning
specific distributional notation, the inverse-gamma density defined
in terms of shape $\alpha$ and rate $\beta$ is denoted by $\mathcal{IG}(\alpha,\beta)$,
the Dirichlet distribution with $k$ concentration parameters by
$\textrm{Dir}(\alpha_{1},\alpha_{2},...,\alpha_{k})$ and the $p$-dimensional
inverse-Wishart distribution with $\nu$ degrees of freedom and scale
matrix $\boldsymbol{\Psi}$ by $\mathcal{IW}_{p}(\nu,\boldsymbol{\Psi})$.

\subsection{Normal regression models}

Here we consider the data set presented in Montgomery et al. (2001,
p.128) concerning 25 direct current (DC) electric charge measurements
(volts) and wind velocity measurements (miles/hour). The goal is to
infer about the effect of wind velocity on the production of electricity
from a water mill. The models under consideration are \\[-2em]
\begin{itemize} 
\item [i)]   $M_{0}$: the null model with the intercept, \\[-1.5em]
\item [ii)]  $M_{1}$: intercept$+(x_{1}-\overline{x}_{1})$, \\[-1.5em]
\item [iii)] $M_{2}$: intercept$+(x_{2}-\overline{x}_{2})$ and \\[-1.5em]
\item [iv)]  $M_{3}$: intercept$+(x_{1}-\overline{x}_{1})+$$x_{1}^{2}$,\\[-1.5em]
\end{itemize} 
where $x_{1}$
is wind velocity and $x_{2}$ is the logarithm of wind velocity. Let
$j$ denote the model indicator, i.e. $j=0,1,2,3$. The likelihood
and prior assumptions are the following 
\begin{eqnarray*}
\mathbf{y}|\boldsymbol{\mathbf{\beta}}_{j},\sigma_{j}^{2}
	&\sim& \mathcal{N}(\mathbf{X}_{j}\boldsymbol{\mathbf{\beta}}_{j},\boldsymbol{I}\sigma_{j}^{2})\\
\mathbf{\boldsymbol{\beta}}_{j}|\sigma_{j}^{2}
	&\sim& \mathcal{N}(\mathbf{0},\mathbf{V}_{j}\sigma_{j}^{2})\\
\sigma_{j}^{2} &\sim& \mathcal{IG}(10^{-3},10^{-3})
\end{eqnarray*}
where $\mathbf{\boldsymbol{\beta}}_{j}$ and $\mathbf{X}_{j}$ correspond
to the regression vector and design matrix of model $j$, respectively,
and $\mathbf{V}_{j}=n^{2}(\mathbf{X}_{j}^{T}\mathbf{X}_{j})^{-1}$
with $n=25$. In relation to the context of Section 2 this is a 2-block
setting where $\boldsymbol{\beta}\equiv\boldsymbol{\theta}$ and $\sigma^{2}\equiv\phib$.
Under this conjugate prior design the distributions $p(\mathbf{\boldsymbol{\beta}}_{j}|\sigma_{j}^{2},\mathbf{y}),\: p(\sigma_{j}^{2}|\mathbf{\boldsymbol{\beta}}_{j},\mathbf{y}),\: p(\mathbf{\boldsymbol{\beta}}_{j}|\mathbf{y})$
and $p(\sigma_{j}^{2}|\mathbf{y})$ are all of known form. 
We treat the posterior distribution as unknown and implement a Gibbs sampler in R. Specifically,
one Gibbs chain is iterated 10,000 times and the first 1000 iterations
are discarded as burn-in, resulting in a final posterior sample of
9,000 draws for each model. 
We calculate two variations of estimator (3), considering: 
i) the true marginals $p(\mathbf{\boldsymbol{\beta}}_{j}|\mathbf{y})$
and $p(\sigma_{j}^{2}|\mathbf{y})$, and ii) Rao-Blackwell estimates
of $p(\mathbf{\boldsymbol{\beta}}_{j}|\mathbf{y})$ and $p(\sigma_{j}^{2}|\mathbf{y})$
based on reduced samples of 200 posterior draws. The two variants
are denoted by $\widehat{m}(\mathbf{y})_{{\scriptscriptstyle \textrm{mp}}}$ and
$\widehat{m}(\mathbf{y})_{{\scriptscriptstyle \textrm{RB}}}$, respectively. 

For comparison reasons, we also consider the following commonly used marginal likelihood estimators: 
the Laplace-Metropolis estimator (Lewis and Raftery, 1997), 
the importance-weighted marginal density estimator of Chen (2005), 
the candidate's estimator from Gibbs sampling (Chib, 1995) and 
the optimal bridge-sampling estimator (Meng and Wong, 1996).
For the Laplace-Metropolis we require only the MCMC estimated 
posterior mean vector and posterior covariance matrix.
For the second estimator, which requires specification of approximating densities, we use normal distributions
for the $\mathbf{\boldsymbol{\beta}}_{j}$'s and inverse gamma distributions
for the $\sigma_{j}^{2}$'s which mimic the respective component-wise
marginal posteriors through moment-fitting. 
In addition, the points which maximize
the unnormalized posterior density of each model are used as posterior
ordinates. 
In order to apply Chib's estimator in a realistic context (using reduced Gibbs sampling) the posterior ordinates (the points maximazing the unnormalized posterior) are decomposed according to the univariate densities. The reduced posterior ordinates are calculated via Rao-Blackwellization based on 9,000 draws from further Gibbs updating (additional sampling is not needed for the simple intercept-mopel).
 Finally, for the optimal bridge-sampling estimator, which is calculated iteratively, 
we utilize the same approximating densities as in the implementation of Chen's estimator 
and iterate 1000 times using the geometric bridge-sampling estimates, also presented in Meng and Wong (1996), as starting values.
The three additional estimators are denoted by $\widehat{m}(\mathbf{y})_{{\scriptscriptstyle \textrm{LM}}}$,
$\widehat{m}(\mathbf{y})_{{\scriptscriptstyle \textrm{Chen}}}$,
$\widehat{m}(\mathbf{y})_{{\scriptscriptstyle \textrm{Chib}}}$ and $\widehat{m}(\mathbf{y})_{{\scriptscriptstyle \textrm{obs}}}$,
respectively.

\begin{table}[h]
\begin{centering}
\begin{tabular}{l@{~}c@{~}cccc}
\hline 
\multirow{2}{*}{\textbf{Estimator}} && \multicolumn{4}{c}{\textbf{Model}}\tabularnewline
\cline{3-6} 
 && $M_0$ & $M_1$ & $M_2$ & $M_3$\tabularnewline
\hline 
Laplace-Metropolis  & $\log\widehat{m}(\mathbf{y})_{{\scriptscriptstyle \textrm{LM}}}$ &-35.1381 & -12.3676  & -0.4044 & -0.9044\tabularnewline
&&(0.0092) &  (0.0124) & (0.0092)& (0.0112)\\[0.5em] 

Importance-weighted & $\log\widehat{m}(\mathbf{y})_{{\scriptscriptstyle \textrm{Chen}}}$&-34.8815 & -13.1407 & -1.5979 & -2.2277\tabularnewline
&&(0.0029) & (0.0039) & (0.0031)& (0.0068)\\[0.5em] 

Candidate's & $\log\widehat{m}(\mathbf{y})_{{\scriptscriptstyle \textrm{Chib}}}$&-34.8789 & -13.1420 & -1.5962 & -2.2337\tabularnewline
&&(0.0020) & (0.0028) & (0.0023)& (0.0067) \\[0.5em] 

Optimal bridge-sampling & $\log\widehat{m}(\mathbf{y})_{{\scriptscriptstyle \textrm{obs}}}$& -34.8807 & -13.1412 & -1.5979 & -2.2294\tabularnewline
&& (0.0011) & (0.0019) & (0.0022)& (0.0030)\\ 

{\underline{Proposed method}} \\ 
\hspace{1em}Exact marginals  
&$\log\widehat{m}(\mathbf{y})_{{\scriptscriptstyle \textrm{mp}}}$ & -34.8786  & -13.1420 & -1.5932 & -2.2302\tabularnewline
&& (0.0023)  & (0.0035) & (0.0030)& (0.0030)\\[0.5em] 

\hspace{1em}Rao-Blackwellization
 &$\log\widehat{m}(\mathbf{y})_{{\scriptscriptstyle \textrm{RB}}}$ & -34.8782  & -13.1405 & -1.5919 & -2.2280\\
&& (0.0023) & (0.0030) & (0.0030) & (0.0033)\\[0.5em]  \hline 
Target value & $\log{m}(\mathbf{y})$ & -34.8797 & -13.1429 & -1.5953 & -2.2270\tabularnewline
\hline 
\end{tabular}
\par\end{centering}

\caption{\label{Tab1}Estimated marginal log-likelihood values compared with true values for Example 1;  
average batch mean estimates (MC errors in parentheses) are presented using 30 batches of size 300. }
\end{table}

In order to calculate MC errors the posterior samples are divided
into 30 batches of 300 draws. Batch mean estimates, MC errors and
the true marginal log-likelihoods are presented in Table \ref{Tab1}. In practical
terms, we found $\widehat{m}(\mathbf{y})_{{\scriptscriptstyle \textrm{LM}}}$ being
the easiest to compute. On the other hand, this estimator performs
poorly in comparison to the others, as seen in Table 1. Variations
of $\widehat{m}(\mathbf{y})_{{\scriptscriptstyle \textrm{LM}}}$ based on multivariate
medians ($L_{1}$ centers) and maximum density points of the unnormalized
posteriors (not presented here) did not yield substantially different estimates.
In contrast, estimators $\widehat{m}(\mathbf{y})_{{\scriptscriptstyle \textrm{Chen}}}$, 
$\widehat{m}(\mathbf{y})_{{\scriptscriptstyle \textrm{Chib}}}$
and $\widehat{m}(\mathbf{y})_{{\scriptscriptstyle \textrm{obs}}}$ perform substantially
better. Implementation for $\widehat{m}(\mathbf{y})_{{\scriptscriptstyle \textrm{obs}}}$
is in general somewhat more complicated in comparison to $\widehat{m}(\mathbf{y})_{{\scriptscriptstyle \textrm{Chen}}}$ as it requires an iterative solution
in addition to specification of approximating densities, whereas $\widehat{m}(\mathbf{y})_{{\scriptscriptstyle \textrm{Chib}}}$ requires additional Gibbs sampling for models $M_1$, $M_2$ and $M_3$. 
The estimators proposed here, $\widehat{m}(\mathbf{y})_{{\scriptscriptstyle \textrm{mp}}}$
and $\widehat{m}(\mathbf{y})_{{\scriptscriptstyle \textrm{RB}}}$, only require
as input the posterior marginal samples and yield comparable batched mean estimates,
with MC errors lower than those of $\widehat{m}(\mathbf{y})_{{\scriptscriptstyle \textrm{Chen}}}$
and just slightly higher than those of estimator $\widehat{m}(\mathbf{y})_{{\scriptscriptstyle \textrm{obs}}}$. 
Also, note that $\widehat{m}(\mathbf{y})_{{\scriptscriptstyle \textrm{Chen}}}$ and $\widehat{m}(\mathbf{y})_{{\scriptscriptstyle \textrm{Chib}}}$ yield higher MC errors for $M_3$.
In addition, the estimates derived through Rao-Blackwellization are
very similar to the estimates obtained from the true marginal
posteriors, while the MC errors are similar across models.

We proceed by testing the ability of the proposed estimator to capture the sensitivity 
of the marginal likelihood over different diffuse prior distributions which have minimal effect on the posterior distributions of 
the regression coefficients $\boldsymbol{\beta}_j$. 
The prior used, with $\mathbf{V}_j=n^{2}(\mathbf{X}_{j}^{T}\mathbf{X}_{j})^{-1}$, corresponds to a Zellner $g$-prior (Zellner, 1986) with $g$ set equal to $n^2$. For this particular data set the posterior distribution of $\boldsymbol{\beta}_j$ is sensitive to the prior when setting $g$ equal to $\sqrt n$ and $n$, which are among the commonly used options (see Fern\'{a}ndez et al., 2001). Therefore, we assume more diffuse priors and use the values of 1000, 1500 and 2000 for $g$; for these choices the posterior distributions are essentially equivalent with posterior expectations (means, standard deviations etc.) being exact up to the 3$\textsuperscript{rd}$ decimal place. We adopt the approach discussed in Section 2.5 using estimator (7) and the model with $g=1000$ as the base model from which we sample from the posterior 10,000 draws discarding the first 1000 as burn-in. Batch mean estimates from the Rao-Blackwell estimator and MC errors, based on 30 batches 300 draws, along with
the true marginal log-likelihoods are presented in Table \ref{Tab2}.

\begin{table}[h!]
\begin{centering}
\begin{tabular}{llcccc}
\hline 
\multirow{2}{*}{$g$\textbf{-prior}} & Target vs.   & \multicolumn{4}{c}{\textbf{Model}}\tabularnewline
\cline{3-6} 
 & Estimate & \multicolumn{1}{c}{$M_0$} & \multicolumn{1}{c}{$M_1$} & \multicolumn{1}{c}{$M_2$} & \multicolumn{1}{c}{$M_3$}\tabularnewline
\hline
\multirow{3}{*}{$g=n^2=625$}  & $\log m(\mathbf{y})$ & -34.8797 & -13.1429 & -1.5953 & -2.2270\tabularnewline
 & $\log\widehat{m}(\mathbf{y})_{{\scriptscriptstyle \textrm{RB}}}$ & -34.8782  & -13.1405 & -1.5919 & -2.2280\tabularnewline
 &  & (0.0023) & (0.0030) & (0.0030) & (0.0033)\tabularnewline
\cline{2-6}
\multirow{3}{*}{$g=1000$}  & $\log m(\mathbf{y})$ & -35.0673  & -13.2125 & -1.0198 & -1.6312\tabularnewline
 & $\log\widehat{m}(\mathbf{y})_{{\scriptscriptstyle \textrm{RB}}}$ & -35.0696 & -13.2063 & -1.0189 & -1.6368\tabularnewline
 &  & (0.0022) & (0.0043) & (0.0024) & (0.0036)\tabularnewline
\cline{2-6}
\multirow{3}{*}{$g=1500$} & $\log m(\mathbf{y})$ & -35.2437 & -13.3897 & -0.8038 & -1.4529\tabularnewline
 & $\log\widehat{m}(\mathbf{y})_{{\scriptscriptstyle \textrm{RB}}}$ & -35.2461 & -13.3836 & -0.8040 & -1.4569\tabularnewline
 &  & (0.0022) & (0.0043) & (0.0032) & (0.0051)\tabularnewline
\cline{2-6}
\multirow{3}{*}{$g=2000$} & $\log m(\mathbf{y})$ & -35.3743 & -13.5616 & -0.7686 & -1.4716\tabularnewline
 & $\log\widehat{m}(\mathbf{y})_{{\scriptscriptstyle \textrm{RB}}}$ & -35.3767 & -13.5556 & -0.7694 & -1.4754\tabularnewline
 &  & (0.0022) & (0.0044)  & (0.0040)  & (0.0067)\tabularnewline
\hline 
\end{tabular}\caption{\label{Tab2}Estimated marginal log-likelihood Rao-Blackwell estimates compared
with the true values for Example 1 for the initial $g$-prior and three diffuse $g$-priors for the
regression vector; average batch mean estimates (MC errors in parentheses)
are presented using 30 batches of size 300. The estimates for $g=1500$ and $g=2000$ are based on posterior samples from the models with $g=1000$.}

\par\end{centering}

\end{table}

As seen in Table \ref{Tab2}, the estimates are accurate despite the fact that the posterior distributions remain the same. In addition, using estimator (7) based on draws from the model with $g=1000$ required only calculation of prior probabilities for the models with $g$ equal to 1500 and 2000, and led to consistent marginal likelihood estimates. The estimates based on the exact marginal posteriors (not presented here) are equivalent.

\subsection{Finite normal mixture models}

In this example we consider the well-known galaxy data which where
initially presented by Postman et al. (1986). The data are velocities
(km's per second) of 82 galaxies from six separated conic sections
of the Corona Borealis region. 
The data set is taken from \texttt{MASS} library in R which contains a ``typo''; the value of the 78\textsuperscript{th} observation was corrected to 26960.
The goal is to investigate whether the galaxies can be classified into different clusters according to
their velocities, as suggested in astronomical theories. Gaussian
finite mixture models are used in the related literature with the
purpose of finding the most plausible number of clusters or components.
Under this modeling assumption, the likelihood of the velocity data
$\mathbf{y}=(y_{1},y_{2},...,y_{n})^{T}$ for a model with $k$ components
$w_{j}\in(0,1)$, such that $\sum_{j}w_{j}=1$ for $j=1,2,...,k$,
is given by 
\begin{equation}
l(\mathbf{y}|\boldsymbol{\mu},\boldsymbol{\sigma}^{2},\mathbf{\boldsymbol{w}})=\underset{i=1}{\overset{n}{\prod}}\,\underset{j=1}{\overset{k}{\sum}}w_{j}\phi(y_{i}|\mu_{j},\sigma_{j}^{2}),
\end{equation}
where $\mathbf{\boldsymbol{w}}=(w_{1},w_{2},...,w_{k})^{T}$,
$\boldsymbol{\mu}=(\mu_{1},\mu_{2},...,\mu_{k})^{T}$, $\boldsymbol{\sigma}^{2}=(\sigma_{1}^{2},\sigma_{2}^{2},...,\sigma_{k}^{2})^{T}$
and $\phi(\cdot)$ is the p.d.f. of the normal distribution. Vectors
$\boldsymbol{\mu}$ and $\boldsymbol{\sigma}^{2}$ consist of the
component-specific means and variances, respectively. As originally
shown in Dempster et al. (1977), any mixture model can be expressed
in terms of missing or latent data; if $z_{i}\in\{1,2,..k\}$ represents
a latent indicator variable associated with observation $y_{i}$,
so that $\Pr(z_{i}=j)=w_{j}$ and $l(y_{i}|z_{i}=j,\mu_{j},\sigma_{j}^{2})=\phi(y_{i}|\mu_{j},\sigma_{j}^{2})$,
then we have that 
\[
l(y_{i},z_{i}=j|\mu_{j},\sigma_{j}^{2})=l(y_{i}|z_{i}=j,\mu_{j},\sigma_{j}^{2})\Pr(z_{i}=j)=\phi(y_{i}|\mu_{j},\sigma_{j}^{2})w_{j}.
\]
Summation over the components $w_{j}$ results in the complete
marginalized data likelihood presented in (10). 

As illustrated in West (1992) and Diebolt and Robert (1994), data-augmentation
facilitates posterior simulation via Gibbs sampling from the full
conditional densities of $\mathbf{\boldsymbol{w}},\boldsymbol{\mu},\boldsymbol{\sigma}^{2}$
and $\mathbf{z}$. The conjugate priors are  $\mu_{j}\sim\mathcal{N}(\mu_{0},\sigma_{0}^{2})$,
$\sigma_{j}^{2}\sim IG(\nu_{0}/2,\delta_{0}/2)$ and $\mathbf{\boldsymbol{w}}\sim\textrm{Dir}(\alpha_{1},\alpha_{2},...,\alpha_{k})$.
The prior for $\mathbf{z}$ is fixed by model design, since $\Pr(z_{i}=j)=w_{j}$.
Gibbs sampling is straightforward to implement, given these prior
assumptions; let $T_{j}=\{i:z_{i}=j\}$ be the set of observation
indices for those $y_{i}$ classified into the $j$-th cluster and
let $n_{j}$ denote the number of observations falling into the $j$-th
cluster. Then, we sample sequentially
\begin{eqnarray*}
z_{i}|\mathbf{y},\mu_{j},\sigma_{j}^{2},w_{j} 
 		&\sim& \Pr(z_{i}=j|\mathbf{y},w_{j},\mu_{j},\sigma_{j}^{2})\propto w_{j}\phi(y_{i}|\mu_{j},\sigma_{j}^{2}),\\
\mu_{j}|\mathbf{y},\sigma_{j}^{2},\mathbf{z} 
		&\sim& \mathcal{N}(\widehat{\mu_{j}},\widehat{s}_{j}^{\,2}),\\
\sigma_{j}^{2}|\mathbf{y},\mu_{j},\mathbf{z} 
		&\sim& \mathcal{IG} \left(  \tfrac{\nu_{0}+n_{j}}{2}, \tfrac{\delta_{0}+\delta_{j}}{2} \right),\\
\boldsymbol{w}|\mathbf{y},\mathbf{z} 
		&\sim& \textrm{Dir}(\alpha_{1}+n_{1},\alpha_{2}+n_{2},...,\alpha_{k}+n_{k}),
\end{eqnarray*}
where $\widehat{\mu_{j}}=\widehat{s}_{j}^{\,2}(\sigma_{0}^{-2}\mu_{0}+\sigma_{j}^{-2}\sum_{i \in T_{j}}y_{i})$,
$\widehat{s}_{j}^{\,2}=(\sigma_{0}^{-2}+\sigma_{j}^{-2}n_{j})^{-1}$ and
$\delta_{j}=\sum_{i \in T_{j}}(y_{i}-\mu_{j})^{2}.$ We are also interested
in models which have a common variance term in (10); in this case the
full conditional of $\sigma^{2}$ is $\mathcal{IG}( \tfrac{\nu_{0}+n}{2}, \tfrac{\delta_{0}+\delta}{2} )$
with $\delta=\sum_{j=1}^{k}\sum_{i \in T_{j}}(y_{i}-\mu_{j})^{2}$. 

A central point in the discussion that follows is the identifiability
problem which is present in mixture models, known as ``label-switching''.
Non-identifiability arises from the fact that relabelling the mixture
components $w_{j}$ does not change the likelihood in (10). Therefore,
when the priors are also invariant to label permutations, the posterior
distribution has $k!$ symmetrical modes. In terms of posterior sampling
this implies that common MCMC samplers will most probably fail to
explore adequately all $k!$ modes as it is very likely that an MCMC
chain will get ``trapped'' in one particular mode thus leaving the
remaining $k!-1$ modes unvisited. A first suggestion proposed in
the early literature is to impose prior ordering constraints, e.g.
$w_{1}<w_{2}<...<w_{k}$ or $\mu_{1} <\mu_{2}<...<\mu_{k}$,
which translate to truncated priors that restrict inference to constrained
unimodal posteriors. Robert and Mengersen (1999) further extended
this strategy to the use of improper priors through reparameterization.
Nevertheless, other authors object to the use of prior identifiability
constraints and recommend sampling from the unconstrained posterior.
Among them, Celeux et al. (2000) propose tempered transition algorithms
and appropriate loss functions for permutation invariant posteriors,
while Marin et al. (2005) suggest ex-post reordering schemes. 

Chib (1995) was the first who estimated directly the marginal likelihoods
of these data for two and three component models via the candidate's formula
and Gibbs updating for the estimation of reduced posterior ordinates.
Nevertheless, as pointed out in Neal (1998), Chib's use of the Gibbs
sampler for mixture models results in biased marginal likelihood estimates
due to lack of label-switching within the Gibbs sampler. A simple
approach to correct for bias is to multiply the marginal likelihood
estimates with a factor of $k!$, but as Neal remarked the bias correction
will only be valid when the symmetrical modes are well-separated (i.e. when label-switching is not likely to occur). 
Therefore, Neal (1998)
suggests either to introduce special relabelling transitions into
the Gibbs sampler or to enforce constrained priors during Gibbs updating
which will be $k!$ times larger than the unconstrained priors, as
general but computationally demanding solutions. Motivated by the
practical bias-correction approach, Berkhof et al. (2003) present
simulation consistent marginal likelihood estimators based on a stratification
principle and ex-post randomly permuted samples. Fr\"{u}hwirth-Schnatter
(2004), on the other hand, recommends to use MCMC samplers which adequately
explore all $k!$ labeling posterior subspaces and presents bridge-sampling
estimators based on draws from the unconstrained random permutation
sampler introduced in Fr\"{u}hwirth-Schnatter (2001). 

We consider the same models as Chib (1995) and show that the estimator
proposed here can accurately estimate the marginal likelihoods either
by taking into account the bias-correction of Neal (1998) or through
the use of MCMC samplers which explore effectively the unconstrained
posterior space. Specifically, interest lies in the 2-component equal-variance
model and 3-component models with equal and unequal variances (i.e.
$k=2,3$), under the prior assumptions $\mu_{0}=20$, $\sigma_{0}^{2}=100$,
$\nu_{0}=6$, $\delta_{0}=40$ and $\alpha_{j}=1$ for $j=1,2,...,k$. 
We further take into account a 4-component equal variance model ($k=4$) with the same prior assumptions. Models with more than four clusters are not considered due to the fact that there is not enough information in the data to support $k>3$, which gives rise to serious convergence problems due to non-identifiability of parameters for more than four clusters; see Carlin and Chib (1995).

The Gibbs sampler for these models can be easily implemented through
package \texttt{bayesmix} (Gr\"{u}n, 2011) in R, which also allows for
ex-post reordering and random permutation sampling. We iterate the
Gibbs sampler 13,000 times and discard the first 1000 iterations as
burn-in. In the context of Section 2 this is a multi-block problem
$(\boldsymbol{\theta}_{1}\equiv\boldsymbol{\mu},\boldsymbol{\theta}_{2}\equiv\boldsymbol{\sigma}^{2},\boldsymbol{\theta}_{3}\equiv\boldsymbol{w})$,
including a latent vector $(\mathbf{u}\equiv\mathbf{z})$ which is
integrated out. We divide the reordered product marginal posterior
sample into $K=30$ batches of $N_K=400$ draws and calculate the marginal likelihood for each batch as 
\[
\widehat{m}(\mathbf{y})=N_{K}^{-1}\underset{n=1}{\overset{N_K}{\sum}}\frac{l(\mathbf{y}|\boldsymbol{\mu}^{(n)},\boldsymbol{\sigma}^{2(n)},\mathbf{\boldsymbol{w}}^{(n)})\pi(\boldsymbol{\mu}^{(n)})\pi(\boldsymbol{\sigma}^{2(n)})\pi(\mathbf{\boldsymbol{w}}^{(n)})}{p(\boldsymbol{\mu}^{(n)}|\mathbf{y})p(\boldsymbol{\sigma}^{2(n)}|\mathbf{y})p(\mathbf{\boldsymbol{w}}^{(n)}|\mathbf{y})}.
\]
Marginal posterior densities are estimated through Rao-Blackwellization
based on reduced samples of size $L=500$, i.e. 
$p(\boldsymbol{\mu}|\mathbf{y})$=$\prod_{j} \big[ \tfrac{1}{L}\sum_{l=1}^{L}p(\mu_{j}|\mathbf{y},\sigma_{j}^{2(l)},\mathbf{z}^{(l)}) \big]$,
$p(\boldsymbol{\sigma}^{2}|\mathbf{y}) 
\\
=\prod_{j}\big[ \tfrac{1}{L} \sum_{l=1}^{L}p(\sigma_{j}^{2}|\mathbf{y},\mu_{j}^{(l)},\mathbf{z}^{(l)}) \big]$
and $p(\boldsymbol{w}|\mathbf{y})=\tfrac{1}{L}\sum_{l=1}^{L}p(\boldsymbol{w}|\mathbf{y},\mathbf{z}^{(l)})$,
for $j=2,3,4$. For the equal-variance models we have that $p(\sigma^{2}|\mathbf{y})=\tfrac{1}{L}\sum_{l=1}^{L}p(\sigma^{2}|\mathbf{y},\mu_{j}^{(l)},\mathbf{z}^{(l)})$.
Table \ref{Tab3} shows batch mean estimates on log scale and the corresponding
MC errors for  the simple estimator $\widehat{m}(\mathbf{y})$, 
the bias-corrected estimator $\widehat{m}_{bc}(\mathbf{y})$ 
(obtained by adding the constant $\log k!$ to $\log\widehat{m}(\mathbf{y})$) and 
the estimator based on ex-post random permutation sampling $\widehat{m}_{rp}(\mathbf{y})$.
\begin{table}[h]
\centering{}%
\begin{tabular}{lcccc}
\hline 
 & \multicolumn{4}{c}{\textbf{Model}}\tabularnewline
\cline{2-5} 
 & 2 clusters & 3 clusters  & 3 clusters & 4 clusters\tabularnewline
{\bf Estimator} & equal & equal & unequal & equal \tabularnewline
 &  variance & variance & variance &  variance\tabularnewline
\hline 
\multirow{2}{*}{$\log\widehat{m}(\mathbf{y})^\star$} 
& -240.458 & -228.597 &-228.595 & -229.027\tabularnewline
& (0.002) & (0.003) & (0.029) & (0.045)\tabularnewline
\multirow{2}{*}{$\log\widehat{m}_{bc}(\mathbf{y})^{\dagger}$} & -239.765 & -226.805 &-226.803 &  -225.849\tabularnewline
& (0.002) & (0.003) & (0.029) & (0.045)\tabularnewline
\multirow{2}{*}{$\log\widehat{m}_{rp}(\mathbf{y})^{\ddagger}$} & -239.762 & -226.778 &-226.771 & -225.922\tabularnewline
& (0.010) & (0.018) & (0.051) & (0.060)\tabularnewline 
\hline
Neal (1998) estimates &-239.764 & -226.803  & -226.791 & -- \\
\hline
\multicolumn{5}{p{12cm}}{ \it \scriptsize 
$^\star\widehat{m}(\mathbf{y})$: Simple (biased) estimator; 
$^\dagger\widehat{m}_{bc}(\mathbf{y})$: Bias-corrected estimator;
$^\ddagger\widehat{m}_{rp}(\mathbf{y})$: Random permutation sampling estimator.} 
\end{tabular}
\caption{\label{Tab3}Estimated marginal log-likelihood values for Example 2;  
average batch mean estimates (MC errors in parentheses) are presented using 30 batches of size 400.}
\end{table}

The benchmark results reported by Neal (1998), based on $10^{8}$
draws from the prior distributions, are also included in Table \ref{Tab3}; the corresponding standard errors are 0.005 for the 2
component equal-variance model, 0.040 for the three component
equal-variance model and 0.089 for the three component unequal-variance
model. It is obvious, that the simple estimator $\widehat{m}(\mathbf{y})$
results in biased estimates, which are very similar to the ones presented
in Chib (1995); see Neal (1998) for the ``typo-corrected'' estimate of
Chib for the $3^{\textrm{rd}}$ model. 
On the other hand, as reflected
in the bias-corrected estimator $\widehat{m}_{bc}(\mathbf{y})$, simply
adding the term $\log k!$ results in accurate marginal likelihood
estimates which are in agreement with the estimates of Neal. 
Interestingly, the MC errors of $\widehat{m}_{bc}(\mathbf{y})$ are similar to the ``coefficients of variation'' in Steele et al. (2006) who handle marginal likelihood estimation through an incremental mixture importance sampling approach based on marginalization. Nevertheless, Steele et al. (2006) adopt different prior assumptions and, therefore, their marginal likelihood estimates are not comparable to the ones in Table 3.  

\begin{figure}[h]
\centering{}\includegraphics[width=\textwidth,height=8cm]{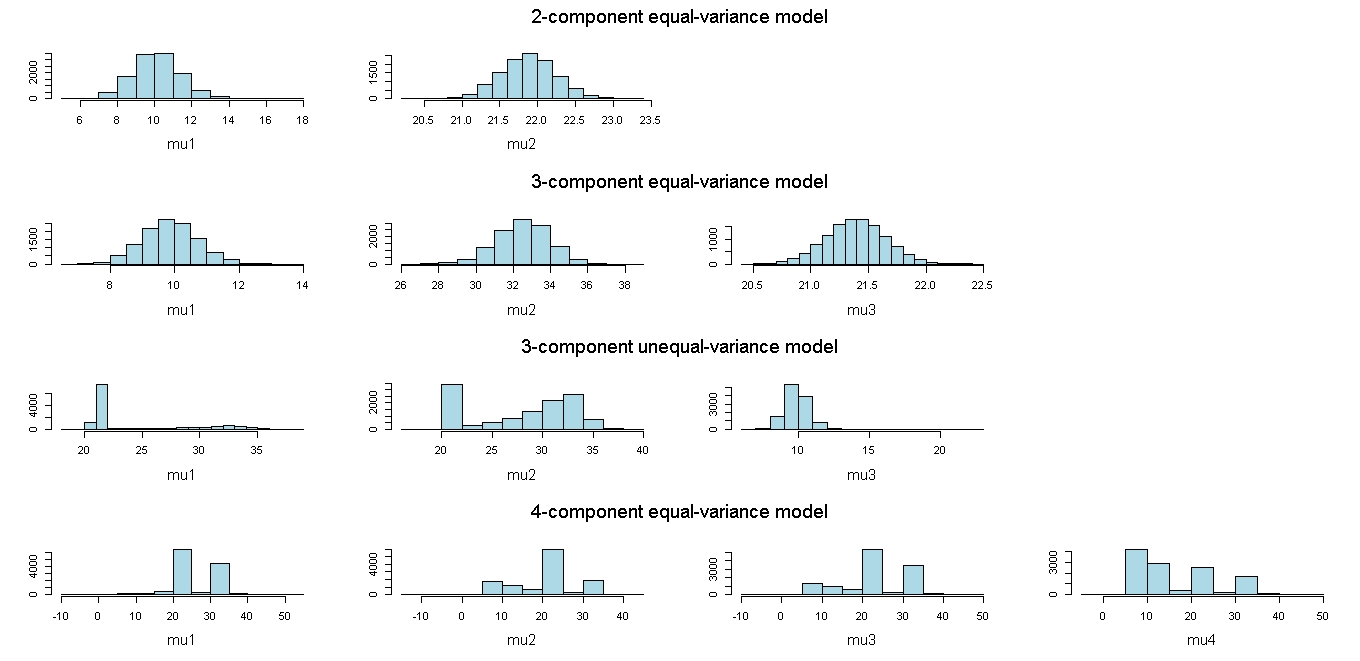}\caption{\label{Fig1}Histograms of posterior means for the three models from Gibbs sampling.}
\end{figure}
\begin{figure}[h]
\centering{}\includegraphics[width=\textwidth,height=8cm]{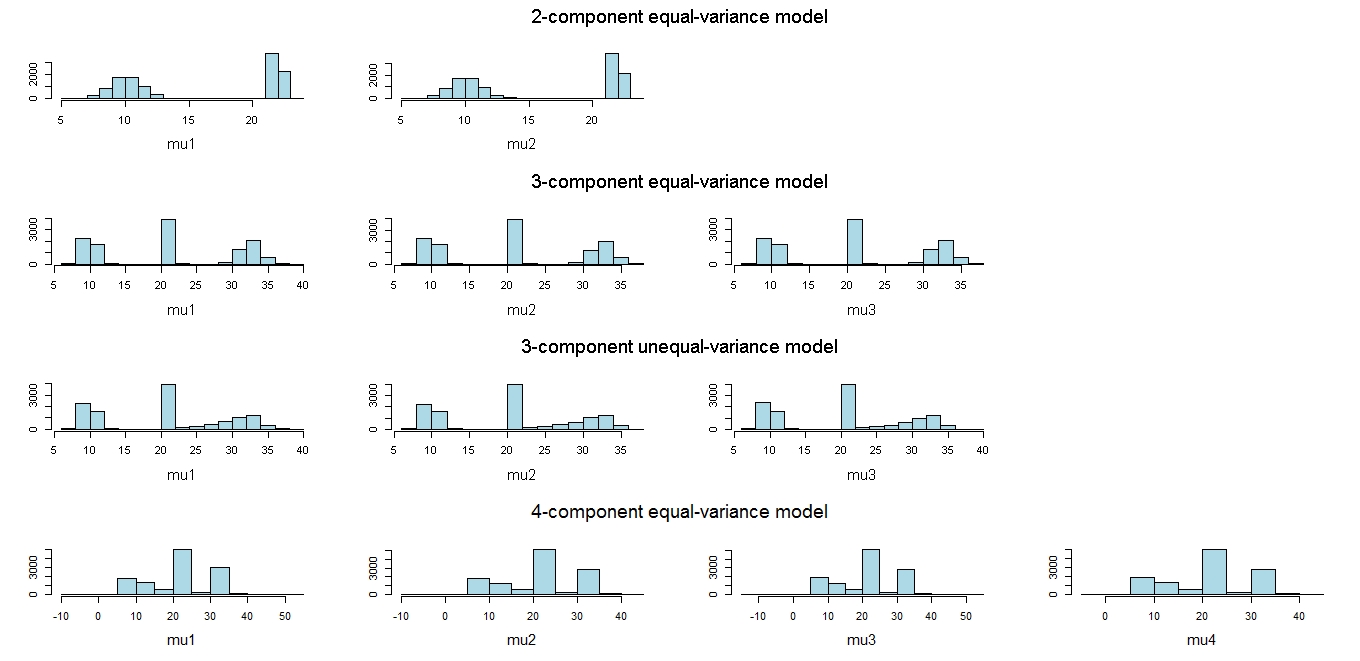}\caption{\label{Fig2}Histograms of posterior means for the three models from random permutation
sampling.}
\end{figure}

Histograms of posterior means from Gibbs sampling are presented in Figure \ref{Fig1}. As seen, the
Gibbs sampler remains in one particular mode for the 2-component and
3-component equal variance models. This is not the case for the 3-component
unequal variance model, where label-switching does actually occur
for parameters $\mu_{1}$ and $\mu_{2}$. 
For the  4-component model label-switching is noticeable for all posterior means. 
Despite that fact, the bias-corrected estimator still performs well for these models. 
Nevertheless, we would not warrant to guarantee that the bias-corrected estimator will always perform well, especially as the number of clusters gets larger and the posterior modes are not well separated.

Alternatively, one can
simply use random permutation sampling and estimate the marginal likelihoods
without the need to account for bias-correction. In addition, random permutation sampling will probably prove to be a more reliable solution for models with many components, since the marginal posteriors from random permutation sampling capture all possible modes. The estimates from
$\widehat{m}_{rp}(\mathbf{y})$ are indeed very similar to Neal's estimates
and to the bias-corrected estimates. The MC errors are slightly higher
for the random permutation estimates, nevertheless, this is understandable
since ex-post random permutation artificially increases MCMC variability.
Histograms of posterior means from random permutation sampling are
presented in Figure \ref{Fig2}; the symmetries in the posterior distributions
due to non-identifiability are now apparent. In accordance to the discussion in Carlin and Chib (1995), the histograms for the 4-component model show that only three modes are estimated efficiently as there is a significant overlap between the 2\textsuperscript{nd} and 3\textsuperscript{rd} mode.

In conclusion, both $\widehat{m}_{bc}(\mathbf{y})$ and $\widehat{m}_{rp}(\mathbf{y})$
yield satisfactory results. For models with a small number of components
(i.e. when label-switching is not likely to occur) the bias-corrected
estimator will most probably be sufficient. For more complicated models
and when the two estimators result in estimates which are in disagreement,
we would recommend to use either the correction for the  candidate estimator proposed by Marin and Robert (2008) 
or the estimator based on alternative MCMC strategies (e.g. Fr\"{u}hwirth-Schnatter, 2001; Geweke, 2007).

\subsection{Longitudinal Poisson models}

As a last example, we consider a data set taken from Diggle et al.
(1995), consisting of seizure counts $y_{it}$ from a group of epilepticts
$(i=1,2,...,59)$ which is monitored initially over an 8-week baseline
period $(t=0)$ and then over four subsequent 2-week periods $(t=1,2,3,4)$.
Each patient is randomly assigned either a placebo or the drug progabide
after the baseline period. This example is chosen mainly because standard
Gibbs sampling is not possible to implement for the model presented
next. In addition, the epilepsy data is also considered by Chib et
al. (1998) and Chib and Jeliazkov (2001) who present marginal likelihood
estimates based on the candidate's formula and Metropolis-Hastings
sampling. Reduced posterior ordinates are calculated through kernel
density estimation in Chib et al. (1998), whereas Chib and Jeliazkov
(2001) employ Metropolis-Hastings updating. For the sake of comparison,
we adopt exactly the same modeling assumptions.

The main model under consideration is
\begin{eqnarray*}
 y_{it}|\boldsymbol{\beta},\mathbf{b}_{i}
 		&\sim& \textrm{Poisson}(\lambda_{it}),\\
 \log\lambda_{it}
 		&=& \log\tau_{it}+\beta_{1}x_{it1}+\beta_{2}x_{it2}+b_{i1}+b_{i2}x_{it2},\\
 \mathbf{b}_{i} &\sim& \mathbf{\mathrm{\mathbf{\mathcal{N}}}}_{2}(\boldsymbol{\eta},\mathbf{D}),
\end{eqnarray*}
where $\tau_{it}$ is the offset which equals 8 when $t=0$ and 2
otherwise, $x_{it1}$ is an indicator of treatment (0 for placebo,
1 for progabide treatment), $x_{it2}$ is an indicator of time period
(0 for baseline, 1 otherwise) and $\mathbf{b}_{i}=(b_{i1},b_{i2})$
are latent random effects for $i=1,2,...,58$ (subject 49 is removed
from the analysis due to unusually high pre-and post-randomization
seizure counts). The prior assumptions are bivariate normal distributions
for $\boldsymbol{\beta}$ and $\boldsymbol{\eta}$ and a bivariate
inverse-Wishart for $\mathbf{D}$, namely $\boldsymbol{\beta}\sim\mathcal{N}_{2}(\mathbf{0},100\boldsymbol{I}_{2})$,
$\boldsymbol{\eta}\sim\mathcal{N}_{2}(\mathbf{0},100\boldsymbol{I}_{2})$
and $\mathbf{D}\sim\mathcal{IW}_{2}(4,\boldsymbol{I}_{2})$, where
$\boldsymbol{I}_{2}$ is the $2\times2$ identity matrix. The full
conditionals of $\boldsymbol{\eta}$ and $\mathbf{D}$ are known,
specifically we have that 
\begin{eqnarray*}
 \boldsymbol{\eta}|\mathbf{D},\mathbf{b}_{1},...,\mathbf{b}_{58}
 			&\sim& \mathcal{N}_{2}(\widehat{\boldsymbol{\eta}},\mathbf{V}),\\ 
\mathbf{D}|\boldsymbol{\eta},\mathbf{b}_{1},...,\mathbf{b}_{58}
 			&\sim& \mathcal{IW}_{2} \Big( 58+4, \, \boldsymbol{I}_{2}+\sum_{i=1}^{58}
 																							(\mathbf{b}_{i}-\boldsymbol{\eta})
 																							(\mathbf{b}_{i}-\boldsymbol{\eta})^{T} \Big),
\end{eqnarray*}
where $\widehat{\boldsymbol{\eta}}=\mathbf{V}\sum_{i=1}^{58}\mathbf{D}^{-1}\mathbf{b}_{i}$
and $\mathbf{V}=(100^{-1}\boldsymbol{I}_{2}+58\mathbf{D}^{-1}){}^{-1}$.
The full conditionals for $\boldsymbol{\beta}$ and the $\mathbf{b}_{i}$'s
are not known distributions and thus standard Gibbs sampling is not
feasible. Another complication is that the integrated sampling likelihood
$l(\mathbf{y}|\boldsymbol{\beta},\boldsymbol{\eta},\mathbf{D})=\prod_{i=1}^{58}\int l(\mathbf{y}_{i}|\boldsymbol{\beta},\mathbf{b}_{i})\pi(\mathbf{b}_{i})d\mathbf{b}_{i}$,
with $l(\mathbf{y}_{i}|\boldsymbol{\beta},\mathbf{b}_{i})=\prod_{t=0}^{4}l(\mathbf{\mathrm{y}}_{it}|\boldsymbol{\beta},\mathbf{b}_{i})$,
is also not available analytically. Therefore, evaluating $l(\mathbf{y}|\boldsymbol{\beta},\boldsymbol{\eta},\mathbf{D})$
requires either numerical integration or some other efficient technique,
such as importance sampling for instance. 

We utilize WinBUGS software (Spiegelhalter et al., 2003) to sample
from the posterior. Specifically, one chain is iterated 31,000 times
and the first 1000 iterations are discarded as burn-in, resulting
in a final sample of 30,000 draws. Posterior means and standard deviations,
for the parameters of scientific interest, are presented in Table
\ref{Tab4}. The estimates for the main model are comparable to the Metropolis-Hastings estimates
presented in Chib et al. (1998). Table \ref{Tab4} also includes the estimates
for the simpler model without the random effects related to time.
For this model we assume a-priori that ${b}_{i} \sim \mathbf{\mathrm{\mathbf{\mathcal{N}}}}({\eta_{1}},{D_{11}})$,
for $i=1,2,...,58$.
In order to keep 
equivalent prior assumptions to the main model, we define the priors as 
$\eta_{1}\sim\mathcal{N}(0,100)$
and $D_{11}\sim\mathcal{IG}(2,1/2)$.

As discussed in Section 2.3, there are two approaches for estimating the marginal
likelihood of this model. The first is to treat
it as a 3-block setting, i.e. consider the product of the marginal
posteriors of $\boldsymbol{\beta},\boldsymbol{\eta},\mathbf{D}$ as
importance sampling density. In this case one needs to estimate the
integrated likelihood which is unknown. The second approach is to
treat the problem as a 4-block setting, i.e. also include the joint marginal
posterior of the $\mathbf{b}_{i}$'s in the importance sampling density.
The advantage with this approach is that we can work directly with
the hierarchical Poisson likelihood. 

Initially, let us consider the
first approach which corresponds to estimator (5) of Section 2.3; in
this case the parameters of scientific interest are $\boldsymbol{\theta}_{1}\equiv\boldsymbol{\beta},\boldsymbol{\theta}_{2}\equiv\boldsymbol{\eta},\boldsymbol{\theta}_{3}\equiv\mathbf{D}$,
while $\mathbf{u}\equiv\{\mathbf{b}_{1},...,\mathbf{b}_{58}\}$ is
used only for Rao-Blackwellization. First, we appropriately re-order
the posterior sample in order to correspond to a sample from the product
marginal posterior and then we split the sample into $K=30$ batches of $N_K=1000$ draws. 
The marginal likelihood estimate for each batch is calculated as 
\[
\widehat{m}_{b_3}(\mathbf{y})=N_{K}^{-1}\underset{n=1}{\overset{N_K}{\sum}}\frac{l(\mathbf{y}|\boldsymbol{\beta}^{(n)},\boldsymbol{\eta}^{(n)},\mathbf{D}^{(n)})\pi(\boldsymbol{\beta}^{(n)})\pi(\boldsymbol{\eta}^{(n)})\pi(\mathbf{D}^{(n)})}{p(\boldsymbol{\beta}^{(n)}|\mathbf{y})p(\boldsymbol{\eta}^{(n)}|\mathbf{y})p(\mathbf{D}^{(n)}|\mathbf{y})}.
\]
Marginal posterior probabilities for $\boldsymbol{\eta}$ and $\mathbf{D}$
are estimated via Rao-Blackwellization based on reduced posterior
samples of $L=200$ draws which are randomly re-sampled from the initial
MCMC sample, i.e. 
$p(\boldsymbol{\eta}|\mathbf{y})$=$\tfrac{1}{L}\sum_{l=1}^{L}p(\boldsymbol{\eta}|\mathbf{D}^{(l)},\mathbf{b}_{1}^{(l)},...,\mathbf{b}_{58}^{(l)})$
and 
$p(\mathbf{D}|\mathbf{y})=\tfrac{1}{L}\sum_{l=1}^{L}p(\mathbf{D}|\boldsymbol{\eta}^{(l)},\mathbf{b}_{1}^{(l)},...,\mathbf{b}_{58}^{(l)})$.
For the marginal posterior of $\boldsymbol{\beta}$ we assume that
$p(\boldsymbol{\beta}|\mathbf{y)}\approx\mathcal{N}_{2}(\tilde{\boldsymbol{\beta}},\tilde{\boldsymbol{\Sigma}})$,
where $\tilde{\boldsymbol{\beta}}$ and $\tilde{\boldsymbol{\Sigma}}$
are estimated from the MCMC output. Similarly to Chib et al. (1998)
and Chib and Jeliazkov (2001), we employ further importance sampling
to evaluate the likelihood $l(\mathbf{y}|\boldsymbol{\beta},\boldsymbol{\eta},\mathbf{D})$.
We employ multivariate normals as importance sampling functions, namely 
$p_{IS}(\mathbf{b}_{1},...,\mathbf{b}_{58}|\mathbf{y})\approx\mathcal{N}_{116}(\tilde{\mathbf{b}},\tilde{\mathbf{B}})$
for the main model and $p_{IS}(b_{1},...,b_{58}|\mathbf{y})\approx\mathcal{N}_{58}(\tilde{\mathbf{b}},\tilde{\mathbf{B}})$
for the reduced model, where $\tilde{\mathbf{b}},\tilde{\mathbf{B}}$
are the vector of means and the complete covariance matrix of the
random effects estimated from the MCMC output. 
Likelihood estimation is based on 100 importance sampling draws.

Based on the alternative 4-block approach, corresponding to estimator (6) of Section 2.3, the batched marginal
likelihood estimates are calculated as
\[
\widehat{m}_{b_4}(\mathbf{y})=N_{K}^{-1}\underset{n=1}{\overset{N_K}{\sum}}\left[\begin{gathered}l(\mathbf{y}|\boldsymbol{\beta}^{(n)},\mathbf{b}_{1}^{(n)},...,\mathbf{b}_{58}^{(n)})\times\\
\pi(\boldsymbol{\beta}^{(n)})\pi(\mathbf{b}_{1}^{(n)},...,\mathbf{b}_{58}^{(n)}|\boldsymbol{\eta}^{(n)},\mathbf{D}^{(n)})\pi(\boldsymbol{\eta}^{(n)})\pi(\mathbf{D}^{(n)})\times\\
\left\{ p(\boldsymbol{\beta}^{(n)}|\mathbf{y})p(\mathbf{b}_{1}^{(n)},...,\mathbf{b}_{58}^{(n)}|\mathbf{y})p(\boldsymbol{\eta}^{(n)}|\mathbf{y})p(\mathbf{D}^{(n)}|\mathbf{y})\right\} ^{-1}
\end{gathered}
\right].
\]
With this approach, there is no need to implement further importance sampling
for evaluating the likelihood function 
since the data conditional on the random effects parameters follow the Poisson distribution. 
Despite this convenient aspect, the downside 
of the 4-block estimator is that it requires estimation of the high-dimensional joint marginal
$p(\mathbf{b}_{1},...,\mathbf{b}_{58}|\mathbf{y})$.
Given that the Rao-Blackwell device cannot be used, 
we adopt a simple
assumption and namely use the importance sampling functions used for likelihood evaluation in estimator $\widehat{m}_{b_3}(\mathbf{y})$, i.e. we assume that $p(\mathbf{b}_{1},...,\mathbf{b}_{58}|\mathbf{y})\approx\mathcal{N}_{116}(\tilde{\mathbf{b}},\tilde{\mathbf{B}})$
for the model with time effects and $p(b_{1},...,b_{58}|\mathbf{y})\approx\mathcal{N}_{58}(\tilde{\mathbf{b}},\tilde{\mathbf{B}})$
for the model not including time effects.

\begin{table}[h]
\begin{centering}
\begin{tabular}{lcccc}
\hline 
\multirow{3}{*}{\textbf{Parameter}} & \multicolumn{2}{c}{\textbf{Model with}} & \multicolumn{2}{c}{\textbf{Model without }}\tabularnewline
 & \multicolumn{2}{c}{\textbf{time effect}} & \multicolumn{2}{c}{\textbf{time effect}}\tabularnewline
\cline{2-5} 
 & Mean & St.Dev. & Mean & St.Dev.\tabularnewline
\hline 
Constant $\eta_{1}$ & 1.065 & 0.146 & 1.095 & 0.138\tabularnewline
Treatment $\beta_{1}$ & -0.0003 & 0.209 & -0.071 & 0.190\tabularnewline
Time $\eta_{2}$ & 0.005 & 0.111 & - & -\tabularnewline
Interaction $\beta_{2}$ & -0.349 & 0.156 & -0.191 & 0.052\tabularnewline
$D_{11}$ & 0.474 & 0.100 & 0.531 & 0.105\tabularnewline
$D_{12}$ & 0.017 & 0.057 & - & -\tabularnewline
$D_{22}$ & 0.243 & 0.063 & - & - \\
\hline 
\\[-0.5em]  
\multicolumn{4}{l}{\underline{Log-marginal likelihood estimates}} \\[0.5em]  
$\log\widehat{m}_{b_3}(\mathbf{y})$ & \multicolumn{2}{c}{-914.992 (0.035)} & \multicolumn{2}{c}{-966.971 (0.018)}\tabularnewline
$\log\widehat{m}_{b_4}(\mathbf{y})$ & \multicolumn{2}{c}{-914.485 (0.137)} & \multicolumn{2}{c}{-966.814 (0.064)}\tabularnewline
\hline 
Chib et al. (1998) &  \multicolumn{2}{c}{-915.404 \hspace{7ex} }  &  \multicolumn{2}{c}{-969.824  \hspace{7ex} } \\ 
Chib \& Jeliazkov (2001) &  \multicolumn{2}{c}{-915.230 \hspace{7ex} }  && \\ 
\hline 
\end{tabular}\caption{\label{Tab4}Posterior means and standard deviations from 30,000 posterior draws
for the parameters of two epilepsy models and average batch mean marginal
likelihood estimates (MC errors in brackets) from 30 batches of size
1000 for the 3-block and 4-block estimators for Example 3.}

\par\end{centering}
\end{table}

Average batch mean marginal likelihood estimates and MC errors 
for the two models in question   
are presented in Table \ref{Tab4}.  
The 3-block approach provides accurate estimates  
with MC errors being very low in comparison to the magnitude of the batched means. 
The 4-block estimates are in agreement with the 3-block estimates, 
but have higher Monte Carlo errors; approximately four times higher than the MC errors of  $\widehat{m}_{b_3}(\mathbf{y})$. 
Nevertheless, this is expected due to the much larger augmented parameter space and 
the use of the normal approximation of the high-dimensional joint posterior of the random effects. 
Overall, estimator $\widehat{m}_{b_4}(\mathbf{y})$ is computationally
less demanding than $\widehat{m}_{b_3}(\mathbf{y})$ 
and the resulting MC errors, although higher than those of $\widehat{m}_{b_3}(\mathbf{y})$, are still relatively low in comparison to the batched mean marginal likelihood values.

 Table \ref{Tab4} also includes the estimates presented in Chib et al. (1998) and Chib and Jeliazkov (2001) from 10,000 posterior draws. 
Both latter studies report standard errors, based on the variance estimator of Newey and West (1987), of approximately 0.1 for the main model. 
The reduced model is briefly considered in Chib et al. (1998) without reporting a standard error. Our 3-block and 4-block estimates are comparable to those of Chib et al. (1998) and Chib and Jeliazkov (2001) but not in strict agreement. Our experience from this particular example is that a long MCMC chain is needed in order to obtain accurate posterior estimates; initial 3-block estimates based on 10,000 posterior draws were actually closer to the estimates of Chib et al. (1998) and Chib and Jeliazkov (2001), namely -915.566 for the main model and -969.580  for the reduced model, but with considerably higher MC errors.

\subsection{MC errors for different number of iterations}

\begin{table}[b!]
\small
\begin{centering}
\begin{tabular}{l@{}lcc}
\hline 
\multirow{2}{*}{\bf Examples} & \multirow{2}{*}{\bf $\:$and models}& \textbf{MCMC} & \textbf{MC} \\ 
                              & & \textbf{length} & \textbf{error}  \\ 
\hline 
&\\[-0.5em] 
\uline{Example 1:}&\uline{~Normal regression models}$^\star$ ($N=9000$) &  &   \\
&\multirow{3}{*}{intercept}                                      & $N$  & 0.0023  \\
                                                               & & $2N$ & 0.0018  \\
                                                               & &      &(0.0016) \\ \cline{2-4}
&\multirow{3}{*}{intercept+wind speed}                           & $N$  & 0.0030  \\
                                                               & & $2N$ & 0.0021  \\
                                                               & &      &(0.0021) \\\cline{2-4}
&\multirow{3}{*}{intercept+$\log$(wind speed)}                   & $N$  & 0.0030  \\
                                                               & & $2N$ & 0.0023  \\ 
                                                               & &      &(0.0021) \\\cline{2-4}
&\multirow{3}{*}{intercept+wind speed+(centered wind speed)$^2$} & $N$  & 0.0033  \\
                                                               & & $2N$ & 0.0026  \\
                                                               & &      &(0.0023) \\\cline{2-4}
&\\[-0.5em] 
\uline{Example 2:}&\uline{~Gaussian mixture models}$^{\dagger}$ ($N=12000$) &  &  \\
&\multirow{3}{*}{2 clusters equal variance}   & $N$  & 0.010  \\
&                                             & $2N$ & 0.007  \\
&                                             &      &(0.007) \\\cline{2-4}
&\multirow{3}{*}{3 clusters equal variance}   & $N$  &0.018   \\
&                                             & $2N$ & 0.015  \\ 
&                                             &      &(0.013) \\\cline{2-4}
&\multirow{3}{*}{3 clusters unequal variance} & $N$  & 0.051  \\
&                                             & $2N$ & 0.039  \\
&                                             &      &(0.036) \\\cline{2-4}
&\multirow{3}{*}{4 clusters equal variance}   & $N$  & 0.060  \\
&                                             & $2N$ & 0.047  \\
&                                             &      &(0.042) \\\cline{2-4}
&\\[-0.5em] 
\uline{Example 3:}&\uline{~Longitudinal Poisson models}$^{\ddagger}$ ($N=30000$) &   &  \\
&\multirow{3}{*}{Model with time effect}    & $N$  & 0.137  \\
&                                           & $2N$ & 0.105  \\  
&                                           &      &(0.097) \\ \cline{2-4}
&\multirow{3}{*}{Model without time effect} & $N$  & 0.063  \\
&                                           & $2N$ & 0.040  \\
&                                           &      &(0.045) \\ 
\hline 
\multicolumn{4}{p{13cm}}{{ \it \scriptsize 
$^\star$MC errors for the Rao-Blackwell estimator for $g=n^2$; $^\dagger$MC errors for the random permutation estimator;
$^\ddagger$MC errors for the 4-block estimator}}
\end{tabular}
\caption{\label{Tab5}MC errors of marginal log-likelihood estimates for MCMC runs of length $N$ and $2N$; the number of batches is 30 for all models. 
The expected MC errors for samples of size $2N$ under the assumption of finite variance are presented in parentheses.}

\par\end{centering}

\end{table}
\normalsize

Here we briefly investigate the issue of finite variance for the illustrated examples presented in this section. 
Following the discussion in Section 2.6, an informal empirical diagnostic for checking whether the variance is finite 
can be performed by comparing the MC errors from MCMC samples of different sizes. 
In general, increasing the MCMC sample by a factor of $d$ should lead to a decrease of MC errors by a factor 
of  $\sqrt d$ for estimators which have finite variance. 
Table \ref{Tab5} depicts the estimated MC errors 
for the Rao-Blackwell estimator for Example 1, the estimator based on random-permutation sampling for Example 2 and the 4-block estimator for Example 3. 
In all cases, the MC errors from the original posterior samples of size $N$ are 
compared with the corresponding errors from samples of size $2N$  
and with the errors from the original samples scaled down by a factor of $\sqrt 2$, 
which are the expected MC errors under the assumption of finite variance. 
From this table, it is evident that the MC errors from the MCMC runs with length equal to $2N$ 
are roughly equal to the MC errors from the chains with length $N$ divided by $\sqrt 2$, for all models, 
indicating that the variance of the corresponding estimators is finite.

\section{Concluding remarks}

In this paper we have presented a method of marginal likelihood estimation based on utilizing the product marginal posterior as importance sampling density. The approach is in general  straightforward to implement even for multi-block parameter
settings as it is non-iterative and does not require adaptations in
MCMC sampling. 
As illustrated, the estimator 
is accurate in capturing changes in the marginal likelihood due to different diffuse prior setups that do not affect the posterior distribution. 
For such cases, the computational demands for estimating marginal likelihoods of competing models under diffuse priors are reduced significantly, since only one MCMC run is required. 
In general, the overall performance of the estimator depends on; i) the efficiency of approximating the joint posterior through independent univariate or multivariate marginals and ii)  the accuracy in estimating marginal posterior densities.

Arguably, the method can fail when the product of marginal posteriors is a poor approximation to the joint posterior. Nevertheless,  appropriate parameter blocking and reparameterizations can always improve the performance of the method, so that it will be feasible to work with a few parameter blocks that are close to orthogonal regardless whether the elements within the blocks are highly correlated. In the three, relatively diverse, examples handled in this paper the natural blocking of the parameters proved to be sufficient in delivering accurate estimates. It is worth noting that similar estimators based on importance sampling from independent posterior factorizations have shown to perform well (Botev et al., 2012; Chan and Eisenstat, 2013). Moreover, independent posterior factorization is also extensively used for the variational Bayes (Bishop, 2006) and expectation-propagation (Minka, 2001) approaches in the maching-learning literature. As a last remark concerning this topic, Ghosh and Clyde (2011) present a methodology for linear and binary regression models that augments non-orthogonal designs to obtain orthogonal designs based on Gibbs sampling for the ``missing'' response variables. With some additional effort one could consider this orthogonalization approach which would guarantee an optimal importance sampling density, thus leaving estimation of univariate marginal posterior densities as the only remaining source of error.

With respect to estimating marginal probabilities, the approach proposed here is particularly suited for Gibbs sampling settings where Rao-Blackwellization can be used to obtain simulation-consistent marginal posterior density estimates. 
Practically, the proposed estimator can get computationally demanding when using Rao-Blackwellization for the entire posterior sample. Nevertheless, the related coding work basically requires averaging and is straighforward, without requiring any special effort in implementation or fine-tuning of parameters in trial and error runs. 
In addition, as illustrated in the examples, the sample needed for Rao-Blackwellization is substantially smaller than the total MCMC sample and is obtained once as a random sub-sample of the MCMC chain.

The approach
can also be applied under other types of MCMC schemes by adopting
other strategies for estimating the marginal posterior densities 
such as normal approximations, fitting posterior moments, kernel methods and so forth. 
In strict theory, the method will not yield unbiased estimates when using such approximating strategies, nevertheless, in practical terms such approaches can often be sufficient and can lead to accurate estimates even for high dimensional multivariate approximations, as demonstrated in Section 3.3. In addition, the degree of bias can be checked indirectly by using the approximating densities as importance sampling densities. It is worth noting, that more elaborated strategies can also be considered, for instance the methods discussed in Oh (1999) based on importance-weighted
marginal density estimation (Chen, 1994) or the integrated nested Laplace approximations (INLA's) presented in Rue et al. (2009).

The advantage of not depending on the type of MCMC scheme used to sample from the posterior becomes obvious for classes of models like the finite normal 
mixtures, considered here, where conventional Gibbs sampling fails to 
explore multi-modal posterior surfaces. This implies that the proposed method
will also work well for models with similar posterior symmetries, 
based on alternative samplers (e.g. Fr\"{u}hwirth-Schnatter, 2001; Geweke, 2007),
without increased complexity in estimation.

A possibly interesting extension
of the idea presented here is to incorporate it within bridge-sampling
estimation by using the product marginal posterior as approximating
density.   

\section*{Acknowledgements}

The authors would like to thank two anonymous referees for their interesting comments and suggestions.

\bibliographystyle{plainnat}

\begin{thebibliography}{22}
\providecommand{\natexlab}[1]{#1}
\providecommand{\url}[1]{\texttt{#1}}
\expandafter\ifx\csname urlstyle\endcsname\relax
  \providecommand{\doi}[1]{doi: #1}\else
  \providecommand{\doi}{doi: \begingroup \urlstyle{rm}\Url}\fi

\bibitem[Ardia Ba\c{s}t\'"{u}rk Hoogerheide and van Dijk(2012)]{1}
Ardia, D., Ba\c{s}t\"{u}rk, N., Hoogerheide, L. and van Dijk, H.K. (2012).
\newblock A comparative study of Monte Carlo methods for efficient evaluation of marginal likelihood.
\newblock\textit{Computational Statistics and Data Analysis}, \textbf{56}, 3398\textendash{}3414.

\bibitem[Berkhof van Mechelen and Gelman(2003)]{2}
Berkhof, J., van Mechelen, I. and Gelman, A. (2003).
\newblock A Bayesian approach to the selection and testing of mixture models.
\newblock\textit{Statistica Sinica}, \textbf{13}, 423\textendash{}442.

\bibitem[Bishop(2006)]{3}
Bishop, C.M. (2006).
\newblock\textit{Pattern Recognition and Machine Learning}. 
\newblock New York: Springer.

\bibitem[Botev L'Ecuyer and Tuffin(2013)]{4}
Botev, Z., L'Ecuyer, P. and Tuffin, B. (2013).
\newblock Markov chain importance sampling with applications to rare event probability estimation.
\newblock\textit{Statistics and Computing}, \textbf{23}, 271\textendash{}285.

\bibitem[Carlin and Chib(1995)]{5}
Carlin, B.P. and Chib, S. (1995). 
\newblock Bayesian model choice
via Markov chain Monte Carlo methods. 
\newblock\textit{Journal of the Royal
Statistical Society B}, \textbf{57}, 473\textendash{}484.

\bibitem[Carlin and Louis(1996)]{6}
Carlin, B.P. and Louis, T. (1996).
\newblock\textit{Bayes and Empirical Bayes Methods for Data Analysis}. 
\newblock London: Chapman \& Hall\textbackslash{}CRC.

\bibitem[Celeux Hurn and Robert(2000)]{7}
Celeux, G., Hurn, M. and Robert, C. (2000). 
\newblock Computational
and inferential difficulties with mixtures posterior distribution. 
\newblock\textit{Journal
of the American Statistical Association}, \textbf{95}, 957\textendash{}979.

\bibitem[Chan and Eisenstat(2013)]{8}
Chan, J. and Eisenstat, E. (2013).
\newblock Marginal likelihood estimation with the cross-entropy method. 
\newblock \textit{Econometric Reviews}, \textit {forthcoming}.

\bibitem[Chen(1994)]{9}
Chen, M.-H. (1994). 
\newblock Importance-weighted marginal
Bayesian posterior density estimation. 
\newblock\textit{Journal of the American
Statistical Association}, \textbf{89}, 818\textendash{}824.

\bibitem[Chen(2005)]{10}
Chen, M.-H. (2005). 
\newblock Computing marginal likelihoods
from a single MCMC output. 
\newblock\textit{Statistica Neerlandica}, \textbf{59}, 16\textendash{}29.

\bibitem[Chib(1995)]{11}
Chib, S. (1995).
\newblock Marginal likelihood from the Gibbs
output. 
\newblock\textit{Journal of the American Statistical Association},
\textbf{90}, 1313\textendash{}132.

\bibitem[Chib Greenberg and Winkelmann(1998)]{12}
Chib, S., Greenberg, E. and Winkelmann, R. (1998).
\newblock Posterior simulation and Bayes factors in panel count data models.
\newblock\textit{Journal of Econometrics}, \textbf{86}, 33\textendash{}54.

\bibitem[Chib and Jeliazkov(2001)]{13}
Chib, S. and Jeliazkov, I. (2001).
\newblock Marginal likelihood
from the Metropolis-Hastings output. 
\newblock\textit{Journal of the American
Statistical Association}, \textbf{96}, 270\textendash{}281.

\bibitem[Dellaportas Forster and Ntzoufras(2002)]{14}
Dellaportas, P., Forster, J.J. and Ntzoufras, I. (2002).
\newblock On Bayesian model and variable selection using MCMC. 
\newblock\textit{Statistics
and Computing}, \textbf{12}, 27\textendash{}36.

\bibitem[Del Moral Doucet and Jasra(2006)]{15}
Del Moral, P., Doucet, A. and Jasra, A. (2006).
\newblock Sequential Monte Carlo samplers.
\newblock \textit{Journal of the Royal Statistical Society B},
\textbf{68}, 411\textendash{}436.

\bibitem[Dempster Laird and Rubin(1977)]{16}
Dempster, A. P., Laird, N. M. and Rubin, D. B. (1977).
\newblock Maximum likelihood from incomplete data via the EM algorithm (with
discussion). 
\newblock\textit{Journal of the Royal Statistical Society B},
\textbf{39}, 1\textendash{}38.

\bibitem[Diebolt and Robert(1994)]{17}
Diebolt, J. and Robert, C.P. (1994).
\newblock Estimation of
finite mixture distributions through Bayesian sampling. 
\newblock\textit{Journal
of the Royal Statistical Society B}, \textbf{56}, 363\textendash{}375.

\bibitem[Diggle Liang and Zeger(1995)]{18}
Diggle, P., Liang, K.-Y. and Zeger, S.L. (1995).
\newblock\textit{Analysis
of Longitudinal Data}. 
\newblock Oxford: Oxford University Press.

\bibitem[Fern\'{a}ndez Ley and Steel(2001)]{19}
 Fern\'{a}ndez, C., Ley, E. and Steel, M.F.J. (2001).
\newblock Benchmark
priors for Bayesian model averaging.
\newblock \textit{Journal of
Econometrics}, \textbf{100}, 381\textendash{}427.

\bibitem[Feroz Balan and Hobson(2011)]{20}
Feroz, F., Balan, S.T. and Hobson, M.P. (2011).
\newblock Bayesian evidence for two companions orbiting HIP 5158. 
\newblock\textit{Monthly Notices of the Royal Astronomical Society}, \textbf{416}, L104\textendash{}L108.

\bibitem[Feroz Hobson and Bridges(2009)]{21}
Feroz, F., Hobson, M.P. and Bridges, M. (2009).
\newblock MULTINEST: an efficient and robust Bayesian inference tool for cosmology and particle physics. 
\newblock\textit{Monthly Notices of the Royal Astronomical Society}, \textbf{398}, 1601\textendash{}1614.

\bibitem[Friel and Pettitt(2008)]{22}
Friel, N. and Pettitt, A.N. (2008).
\newblock Marginal likelihood
estimation via power posteriors. 
\newblock\textit{Journal of the Royal Statistical
Society B}, \textbf{70}, 589\textendash{}607.

\bibitem[Friel and Wyse(2012)]{23}
Friel, N. and Wyse, J. (2012).
\newblock Estimating the evidence -- a review. 
\newblock\textit{Statistica Neerlandica}, \textbf{66}, 288\textendash{}308.

\bibitem[Fr\"{u}hwirth-Schnatter(2001)]{24}
Fr\"{u}hwirth-Schnatter, S. (2001).
\newblock Markov chain Monte
Carlo estimation of classical and dynamic switching and mixture models.
\newblock\textit{Journal of the American Statistical Association}, \textbf{96}, 194\textendash{}209.

\bibitem[Fr\"{u}hwirth-Schnatter(2004)]{25}
Fr\"{u}hwirth-Schnatter, S. (2004).
\newblock Estimating marginal
likelihoods for mixture and Markov switching models using bridge sampling
techniques. 
\newblock\textit{The Econometrics Journal}, \textbf{7}, 143\textendash{}167.

\bibitem[Gelfand and Smith(1990)]{26}
Gelfand, A.E. and Smith, A.F.M. (1990).
\newblock Sampling-based
approaches to calculating marginal densities. 
\newblock\textit{Journal of the
American Statistical Association}, \textbf{85}, 398\textendash{}409.

\bibitem[Gelman and Rubin(1992)]{27}
Gelman, A. and Rubin, D.B. (1992).
\newblock Inference from iterative simulation
using multiple sequences. 
\newblock\textit{Statistical Science}, \textbf{7}, 457\textendash{}511.

\bibitem[Geweke(2007)]{28}
Geweke, J. (2007).
\newblock Interpretation and inference in
mixture models: simple MCMC works. 
\newblock\textit{Computational Statistics
and Data Analysis}, \textbf{51}, 3529\textendash{}3550. 

\bibitem[Geyer(1992)]{29}
Geyer, C.J. (1992).
\newblock Practical Markov chain Monte Carlo. 
\newblock\textit{Statistical Science}, \textbf{4}, 473\textendash{}483.

\bibitem[Ghosh and Clyde(2011)]{30}
Ghosh, J. and Clyde, M.A. (2011).
\newblock Rao-Blackwellization for Bayesian variable selection and model averaging in linear and binary regression: a novel data augmentation approach.
\newblock \textit{Journal of the American Statistical Association}, \textbf{106}, 1041\textendash{}1052.

\bibitem[Gilks and Roberts(1996)] {31}
Gilks, W.R. and Roberts, G.O. (1996). 
\newblock Strategies for improving MCMC.
\newblock In \textit{Markov Chain Monte Carlo in Practice}, 6, eds. W.R. Gilks, S. Richardson and D.J. Spiegelhalter, London: Chapman \& Hall\textbackslash{}CRC,
pp. 89\textendash{}114.

\bibitem[Green(1995)]{32}
Green, P.J. (1995).
\newblock Reversible jump Markov chain Monte
Carlo computation and Bayesian model determination. 
\newblock\textit{Biometrika},
\textbf{82}, 711\textendash{}732.

\bibitem[Gr\"{u}n(2011)]{33}
Gr\"{u}n, B. (2011).
\newblock\textit{bayesmix: Bayesian Mixture
Models with JAGS. }
\newblock Available at http://cran.r-project.org/web/packages/bayesmix/index.html.

\bibitem[Kass and Raftery(1995)]{34}
Kass, R.E. and Raftery, A.E. (1995). 
\newblock Bayes factors and model uncertainty. 
\newblock\textit{Journal of the American Statistical
Association}, \textbf{90}, 773\textendash{}795.

\bibitem[Lewis and Raftery(1997)]{35}
Lewis, S.M. and Raftery, A.E. (1997). 
\newblock Estimating Bayes
factors via posterior simulation with the Laplace-Metropolis estimator.
\newblock\textit{Journal of the American Statistical Association}, \textbf{92}, 648\textendash{}655.

\bibitem[Marin Mengersen and Robert(2005)]{36}
Marin, J.-M., Mengersen, K. and Robert, C. (2005).
\newblock Bayesian modelling and inference on mixtures of distributions. 
\newblock In
\textit{Handbook of Statistics}, vol. 25, eds. C. Rao and D. Dey, New
York: Elsevier, pp. 459\textendash{}507. 

\bibitem[Marin and Robert(2008)]{36b}
\newblock Marin, J.-M and Robert, C.P. (2008).  
\newblock Approximating the marginal likelihood in mixture models. 
\textit{ Bulletin of the Indian Chapter of ISBA}, {\bf V}(1), 2--7; 
also available as {\tt arXiv0804.2414}. 

\bibitem[Meng and Wong(1996)]{37}
Meng, X.-L. and Wong, W.H. (1996). 
\newblock Simulating ratios
of normalizing constants via a simple identity: a theoretical exploration.
\newblock\textit{Statistica Sinica}, \textbf{6}, 831\textendash{}860.

\bibitem[Minka(2001)]{38}
Minka, T.P. (2001). 
\newblock Expectation propagation for approximate Bayesian inference.
\newblock\textit{Uncertainty in Artificial Intelligence}, \textbf{17}, 362\textendash{}369.

\bibitem[Montgomery Peck and Vining(2001)]{39}
Montgomery, D.C., Peck, E.A. and Vining, G.G. (2001).
\newblock\textit{Introduction to Linear Regression Analysis}. 
\newblock New York: John
Wiley.

\bibitem[Neal(1998)]{40}
Neal, R.M. (1998). 
\newblock Erroneous results in \textquoteleft{}Marginal
likelihood from the Gibbs output\textquoteright{}. 
\newblock Available at http://www.cs.utoronto.ca/radford/radford@stat.utoronto.ca.

\bibitem[Neal(2001)]{41}
Neal, R.M. (2001).
\newblock Annealed importance sampling. 
\newblock\textit{Statistics and Computing}, \textbf{11}, 125\textendash{}139.

\bibitem[Newey and West(1987)]{42}
Newey, W. K. and West, K. D. (1987).
\newblock A simple positive semi-definite heteroskedasticity and autocorrelation consistent covariance matrix. 
\newblock \textit{Econometrica}, \textbf{55}, 703\textendash{}708.

\bibitem[Newton and Raftery(1994)]{43}
Newton, M.A. and Raftery, A.E. (1994).
\newblock Approximate
Bayesian inference with the weighted likelihood bootstrap. 
\newblock\textit{Journal
of the Royal Statistical Society B}, \textbf{56}, 3\textendash{}48.

\bibitem[Ntzoufras, Katsis and Karlis(2005)]{44} 
Ntzoufras, I., Katsis, A. and Karlis, D. (2005). 
\newblock Bayesian assessment of the distribution of insurance claim counts using reversible jump MCMC. 
\newblock \textit{North American Actuarial Journal}, \textbf{9}, 90\textendash{}108.

\bibitem[Oh(1999)]{45}
Oh, M.-S. (1999).
\newblock Estimation of posterior density
functions from a posterior sample. 
\newblock\textit{Computational Statistics
and Data Analysis}, \textbf{29}, 411\textendash{}427.

\bibitem[Parise and Welling(2007)] {46}
Parise, S. and Welling, M. (2007). 
\newblock Bayesian model scoring in Markov random fields.
\newblock In \textit{Advances in Neural Information Processing Systems}, 19, eds. B. Sch\"{o}lkopf, J. Platt and T. Hoffman, Cambridge, MA: MIT Press,
pp. 1073\textendash{}1080.


\bibitem[Postman Huchra and Geller(1986)]{47}
Postman, M., Huchra, J.P. and Geller, M.J. (1986).
\newblock Probes of large-scale structures in the Corona Borealis region. 
\newblock\textit{The
Astronomical Journal}, \textbf{92}, 1238\textendash{}1247.

\bibitem[Raftery Newton Satagopan and Krivitsky(2007)] {48}
Raftery, A.E., Newton, M.A., Satagopan, J.M. and Krivitsky, P.N. (2007). 
\newblock Estimating the integrated likelihood via posterior simulation using the harmonic mean identity.
\newblock In \textit{Bayesian Statistics}, 8, eds. J.M. Bernando, M.J. Bayarri, J.O. Berger, A.P. Dawid, D. Heckerman, A.F.M. Smith and M. West, Oxford, U.K.: Oxford University Press,
pp. 1\textendash{}45.

\bibitem[Robert and Mengersen(1999)]{49}
Robert, C. P. and Mengersen, K. L. (1999).
\newblock Reparameterization
issues in mixture modelling and their bearing on MCMC algorithms.
\newblock\textit{Computational Statistics and Data Analysis}, \textbf{29}, 325\textendash{}343.

\bibitem[Rue, Martino and Chopin(2009)]{50}
Rue, H., Martino, S. and Chopin, N. (2009).
\newblock Approximate Bayesian inference for latent Gaussian models by using integrated nested Laplace approximations (with discussion).
\newblock \textit{Journal of the Royal Statistical Society B}, \textbf{71}, 319\textendash{}92.

\bibitem[Scott, D.W. (1992)]{51}
Scott, D.W. (1992). 
\newblock\textit{Multivariate Density
Estimation}. 
\newblock New York: John Wiley.

\bibitem[Skilling(2006)]{52}
Skilling, J. (2006). 
\newblock Nested sampling for general Bayesian computation.
\newblock\textit{Bayesian Analysis}, \textbf{1}, 833\textendash{}860.

\bibitem[Spiegelhalter Thomas Best and Lunn(2003)]{53}
Spiegelhalter, D., Thomas, A., Best, N. and Lunn,
D. (2003). 
\newblock\textit{WinBUGS User Manual, Version 1.4}. 
\newblock UK: MRC Biostatistics
Unit, Institute of Public Health and Department of Epidemiology and
Public Health, Imperial College School of Medicine. Available at http://www.mrc-bsu.cam.ac.uk/bugs/winbugs/manual14.pdf 

\bibitem[Steele Raftery and Emond(2006)]{54}
Steele, R.J., Raftery, A.E. and Emond, M.J. (2006). 
\newblock Computing normalizing constants for finite mixture models via incremental mixture importance sampling.
\newblock\textit{Journal of Computational and Graphical Statistics}, \textbf{15}, 712\textendash{}734.

\bibitem[Tanner and Wong(1987)]{55}
Tanner, M.A. and Wong, W. (1987). 
\newblock The calculation
of posterior distributions by data augmentation (with discussion).
\newblock\textit{Journal of the American Statistical Association}, \textbf{82}, 528\textendash{}550.

\bibitem[Vitoratou {\it et al.} (2013)]{56b}
Vitoratou, S., Ntzoufras, I. and Moustaki I. (2013).
\newblock Explaining the behavior of joint and marginal Monte Carlo estimators in latent variable models with independence assumptions. 
\newblock{\tt arXiv:{\bf 1311.0656}[stat.CO]} (submitted).



\bibitem[Weinberg(2012)]{56}
Weinberg, M.D. (2012). 
\newblock Computing the Bayes factor
from a Markov chain Monte Carlo simulation of the posterior distribution.
\newblock\textit{Bayesian Analysis}, \textbf{7}, 737\textendash{}770.

\bibitem[West(1992)] {57}
West, M. (1992). 
\newblock Modelling with mixtures (with discussion).
\newblock In \textit{Bayesian Statistics}, 4, eds. J.M. Bernando, J.O. Berger,
A.P. Dawid and A.F.M. Smith, Oxford, U.K.: Oxford University Press,
pp. 503\textendash{}524.

\bibitem[Zellner(1986)]{58}
 Zellner, A. (1986).
\newblock On assessing prior distributions and Bayesian
regression analysis with g-prior distributions.
\newblock In \textit{Bayesian Inference and Decision Techniques:
Essays in Honour of Bruno de Finetti}, eds. P. Goel and A.
Zellner, Amsterdam: North-Holland, pp. 233\textendash{}243. 

\end{thebibliography}

\end{document}